%% file: main.tex
\documentclass[sigplan,nonacm]{acmart}

\usepackage{multirow}
\usepackage{xspace}
\usepackage{graphicx}      
\usepackage{subcaption}    
\usepackage{enumitem}
\setlist[itemize]{leftmargin=*}
\setlist[enumerate]{leftmargin=*}

\input{command.tex}
\begin{document}

\title{Accelerating Mixture-of-Experts Inference by Hiding Offloading Latency with Speculative Decoding}

\author{Zhibin Wang}
\affiliation{%
  \institution{State Key Laboratory for Novel Software Technology, Nanjing University}
  \city{Nanjing}
  \country{China}
}
\author{Zhonghui Zhang}
\affiliation{%
  \institution{State Key Laboratory for Novel Software Technology, Nanjing University}
  \city{Nanjing}
  \country{China}
}
\author{Yuhang Zhou}
\affiliation{%
  \institution{State Key Laboratory for Novel Software Technology, Nanjing University}
  \city{Nanjing}
  \country{China}
}
\author{Zibo Wang}
\affiliation{%
  \institution{State Key Laboratory for Novel Software Technology, Nanjing University}
  \city{Nanjing}
  \country{China}
}
\author{Mo Zhou}
\affiliation{%
  \institution{State Key Laboratory for Novel Software Technology, Nanjing University}
  \city{Nanjing}
  \country{China}
}
\author{Peng Jiang}
\affiliation{%
  \institution{State Key Laboratory for Novel Software Technology, Nanjing University}
  \city{Nanjing}
  \country{China}
}
\author{Weilin Cai}
\affiliation{%
  \institution{The Hong Kong University of Science and Technology (Guangzhou)}
  \city{Guangzhou}
  \country{China}
}
\author{Chengying Huan}
\affiliation{%
  \institution{State Key Laboratory for Novel Software Technology, Nanjing University}
  \city{Nanjing}
  \country{China}
}
\author{Rong Gu}
\affiliation{%
  \institution{State Key Laboratory for Novel Software Technology, Nanjing University}
  \city{Nanjing}
  \country{China}
}
\author{Sheng Zhong}
\affiliation{%
  \institution{State Key Laboratory for Novel Software Technology, Nanjing University}
  \city{Nanjing}
  \country{China}
}
\author{Chen Tian}
\affiliation{%
  \institution{State Key Laboratory for Novel Software Technology, Nanjing University}
  \city{Nanjing}
  \country{China}
}

\renewcommand{\shorttitle}{}
\renewcommand{\shortauthors}{}

\begin{abstract}
    Recent advancements in Mixture of Experts (MoE) models have significantly increased their parameter scale as well as model performance. Extensive offloading techniques have been proposed to address the GPU memory limitations of MoE inference. However, due to the I/O bottleneck and sparse computation of MoE models, existing offloading techniques still suffer from low hardware utilization. To fully utilize the hardware resources, we propose SpecMoEOff, which employs the speculative decoding technique to enlarge the workload of each expert. SpecMoEOff orchestrates the GPU and CPU by both theoretical and empirical roofline analysis. In addition, we develop a dedicated CPU chunked attention verification kernel to fit the speculative decoding in offloading scenarios as well as minimizing the additional overhead led by draft models. SpecMoEOff further integrates an optimizer to automatically tune the hyperparameters of speculative decoding under given hardware and workload. Experimental results show that SpecMoEOff achieves up to 2.5$\times$ decode throughput improvement over the state-of-the-art MoE offloading techniques.
\end{abstract}

\maketitle 

\section{Introduction}

Mixture of Experts (MoE) models have shown significant improvements in performance for various machine learning tasks, as evident by recent models like Qwen~\cite{qwen2025qwen25technicalreport}, Gemini~\cite{google_gemini_website}, DeepSeek~\cite{deepseek_website}, Mixtral~\cite{jiang2024mixtral}, LLaMA~\cite{meta_llama_website}, and others.
MoE models utilize a sparse architecture where only a subset of experts is activated for each input, allowing for larger models without a proportional increase in computational cost.
Recently, mainstream MoE models reached a scale of trillion parameters, e.g., DeepSeek-R1~\cite{deepseek_website}, which needs 671 GB of memory to store the model parameters even with 8-bit quantization.
The memory footprint of MoE models poses a significant challenge for deployment on consumer-grade GPUs, which typically have limited memory capacity.

\begin{figure}[tbp]
    \centering
    \includegraphics[width=\linewidth]{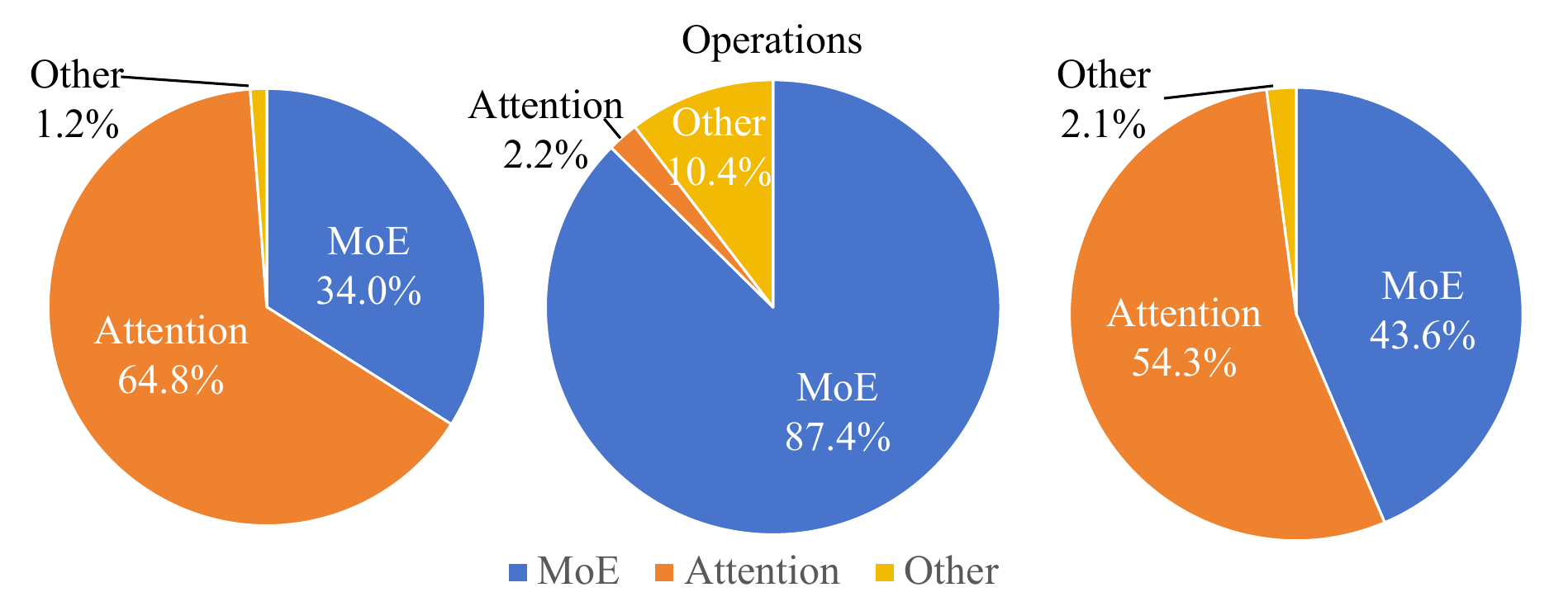}
    \caption{Proportion of cost of three kinds of layers.}
    \label{fig:pie}
\end{figure}

Offloading~\cite{sheng2023flexgen,deepspeedinference}, a technique leveraging the memory of external devices, has been proposed to address the limitations of GPU memory for handling large models. This approach involves transferring parts of the model or intermediate data to external memory (e.g., CPU RAM or disk storage) during inference, allowing the GPU to focus on the most critical computations. Regarding MoE models, offloading can be applied to 1) the \emph{model parameters}~\cite{cao2025moe,yu2025fmoefinegrainedexpertoffloading,sheng2023flexgen}, mainly the expert weights, which constitute a significant portion (e.g., 97\% in Mixtral 8x7B) of the model's memory footprint, and 2) the \emph{KV cache}~\cite{sheng2023flexgen,kwon2023efficient,cao2025moe}, which is used to store the key and value pairs for attention mechanisms, can become exceedingly large in scenarios with large batch sizes, as shown in Figure~\ref{fig:pie}. In addition to the memory, computation of the data can also be offloaded to external devices, such as CPUs, to mitigate the high transfer cost of data between the GPU and external memory.

One main streaming approach~\cite {rajbhandari2022deepspeedmoeadvancingmixtureofexpertsinference,yu2025fmoefinegrainedexpertoffloading} for offloading MoE models is to leverage the sparsity of MoE models under small batch sizes, typically batch size 1, to minimize the transfer cost between the GPU HBM and CPU DRAM by only transferring the weights of the activated experts to the GPU HBM. By further incorporating expert prediction and caching frequently activated (hot) experts, this approach achieves over 80\% reduction in the transfer cost of expert weights and significantly improves latency.
However, loading one expert can only process one or a few inputs, which makes transfer costs become the bottleneck, resulting in only 0.76\% GPU utilization in Mixtral 8x7B model decoding on A30 GPU, which can be obtained by the model in Section~\ref{sec:roofline_visual}.

Another throughput-oriented approach, MoE-Lightning~\cite{cao2025moe}, combats the sparsity of MoE layers by increasing the batch size.
Once expert weights are loaded into HBM, the GPU can process multiple inputs simultaneously, yielding much higher throughput than the small-batch approach.
To handle large batch sizes, MoE-Lightning maintains the KV cache in the CPU DRAM and performs the attention computation on the CPU.
Unfortunately, due to the limited DRAM and large KV cache consumption of each request, the batch size is still limited. As a result, MoE-Lightning only achieves a theoretical GPU utilization of 3.13\%, as revealed in Section~\ref{sec:roofline_visual}.

In this paper, we propose \name, a novel MoE offloading system that, to the best of our knowledge, is the first to employ speculative decoding for maximizing GPU utilization under the MoE offloading scenario. Speculative decoding~\cite{xia2022speculative} is proposed to improve the parallelism and reduce the memory access of KV cache in GPU-based inference of large language models. It adopts a lightweight \emph{draft model} to generate multiple draft tokens, which are then verified by the original \emph{target model} once, thereby increasing the process tokens in each forward pass. We observe that such an increased workload naturally aligns with the MoE offloading setting, where GPU resources are often underutilized.
\begin{figure*}[t]
    \centering
    \includegraphics[width=\linewidth]{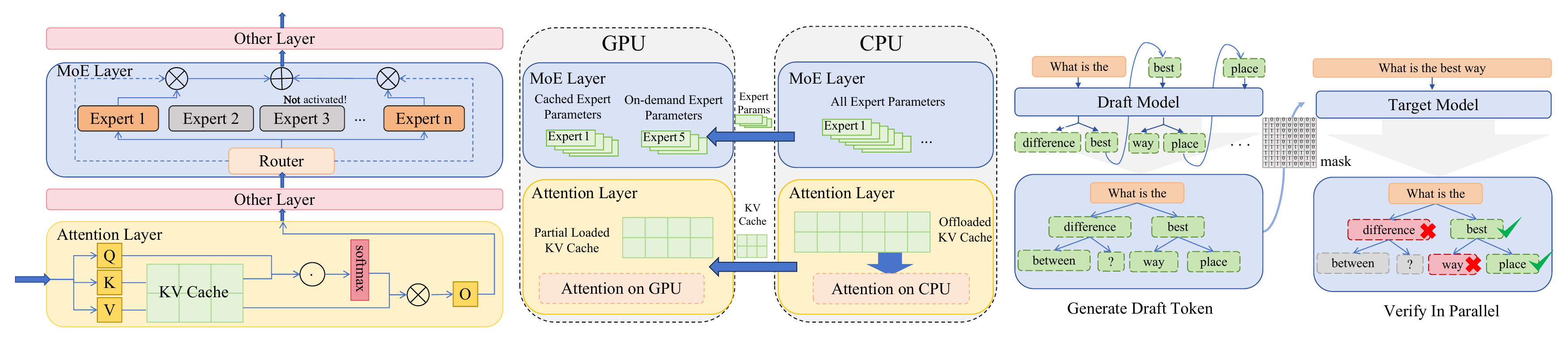}
    \vspace{-0.3in}
    \caption{MoE model, offloading, and speculative decoding.}
    \label{fig:background}
\end{figure*}

However, employing speculative decoding in MoE models is non-trivial as the following issues need to be addressed:
\begin{itemize}
    \item \emph{Issue 1: Inefficient speculative decoding verification.} The performance of speculative decoding verification relies on efficient chunked attention mechanisms. However, existing implementations for GPU require frequent offloading of KV cache to CPU memory, incurring expensive CPU-GPU data transfers. Meanwhile, current CPU chunked attention only supports either full attention or single-token attention and lacks optimizations for speculative decoding.
    \item \emph{Issue 2: Discrepancy between draft and target models.} As discussed in Section~\ref{sec:memory_conscious_draft_model_execution}, the draft model may also exceed GPU HBM capacity and require offloading. In each speculative decoding iteration, the draft model is invoked multiple times, whereas the target model is called only once. This discrepancy makes it non-trivial to directly extend target model offloading strategies to the draft model.
    \item \emph{Issue 3: Complexity of hyperparameter tuning.} Beyond the hyperparameters in existing offloading systems, speculative decoding introduces additional parameters such as the number of draft tokens. The interplay between these hyperparameters further complicates the optimization.
\end{itemize}

Accordingly, the contributions of this paper are as follows:
\begin{itemize}
    \item In Section~\ref{sec:roofline_analysis}, we conduct a roofline analysis of existing MoE offloading techniques, illustrating that memory access and transfer are still the common bottleneck, thereby motivating the need for speculative decoding.
    \item In Section~\ref{sec:overview}, we orchestrate the CPU and GPU execution of speculative decoding in MoE offloading scenarios, by combining the existing offloading techniques and our proposed optimizations.
    \item In Section~\ref{sec:chunked_attention} (corresponding to issue 1), we develop a CPU-oriented chunked attention verification kernel, which is designed to efficiently verify the draft tokens in speculative decoding for CPU execution by leveraging the Intel MKL to accelerate matrix multiplication and reducing the memory consumption of the mask.
    \item In Section~\ref{sec:memory_conscious_draft_model_execution} (corresponding to issue 2), we propose a draft model offloading strategy that considers the discrepancy between the draft and target models to prioritize the caching of the draft model and CPU/GPU separation for better concurrency.
    \item In Section~\ref{sec:hyperparameter_optimizer} (corresponding to issue 3), we develop a hyperparameter optimizer that combines the theoretical convex optimization and empirical profiling-based estimation to automatically tune the hyperparameters of \name under given hardware, model, and workload.
    \item In Section~\ref{sec:evaluation}, we evaluate \name on various workloads and hardware configurations, demonstrating that \name achieves $2.5\times$ average speedup over the state-of-the-art MoE offloading systems.
\end{itemize}


\section{Background}
\label{sec:background}

\subsection{Mixture of Experts (MoE) Models}
The Mixture of Experts (MoE)~\cite{shazeer2017outrageously} architecture has been widely adopted by recent state-of-the-art models such as Qwen~\cite{qwen2025qwen25technicalreport}, Gemini~\cite{google_gemini_website}, deepseek~\cite{deepseek_website}, mixtral~\cite{jiang2024mixtral} and LLaMA~\cite{meta_llama_website}.
As shown in Figure~\ref{fig:background}(a), MoE models mainly include following components:
\begin{itemize}
    \item \textbf{Attention Layer:} Attention mechanism is a core component of transformer-based architectures, enabling the model to weigh the relevance of different input tokens when generating outputs. In inference, especially for decoding stage, the attention operation is dominated by the access of the KV cache, which stores the key and value pairs for all previous tokens.
    \item \textbf{MoE Layer:} The MoE layer replaces the standard FFN with a mixture of experts, allowing the model to dynamically select a subset of experts for each input token. Though the theoretical computation of the MoE layer is similar to the standard FFN, it reduces the arithmetic intensity, i.e., the ratio of computation to memory access, due to the sparse activation of experts.
    \item \textbf{Other Layers:} Besides the attention and MoE layers, the model also includes other standard transformer components such as layer normalization, residual connections, and embedding layers. These components are essential for the overall functionality of the model but typically have a smaller impact on memory and computation compared to the attention and MoE layers.
\end{itemize}

Figure~\ref{fig:pie} demonstrates the proportion of memory and computation cost of the three kinds of layers in the Mixtral-8x7B model when running the APPS dataset on an A30 GPU with 250 GB of CPU memory.
We observe that the KV cache and weight of experts dominate the memory consumption as well as the memory access cost of MoE models. Regarding computation operations, the MoE layer dominates while the attention layer is negligible. This divergence in memory and computation characteristics leads to different offloading strategies for MoE layers and attention layers as we will discuss in next section.

\subsection{Offloading techniques}

\stitle{Traditional tensor-based offloading.} In the age of traditional transformer-based LLMs (using FFN instead of MoE), researchers have proposed various offloading techniques~\cite{ren2021zerooffloaddemocratizingbillionscalemodel,sheng2023flexgen,aminabadi2022deepspeedinferenceenablingefficient} to address the memory limitations of GPU. One of the representative works is FlexGen~\cite{sheng2023flexgen}, which considers the partitioning of the tensors and organizes the order of the memory access to minimize the transfer cost. This becomes a standard offloading technique for LLMs, which can be applied to MoE models as well, while \name also adopts this technique.

\stitle{Offloading for MoE layers.} When shifting the focus from FFN to MoE layers, the sparse activation of experts introduces new challenges and opportunities for offloading. One branch of approach is to exploit the sparsity of expert activations by only loading and caching the weights of the activated experts, thereby reducing the memory footprint and transfer cost. This works for small batch size, especially for batch size 1, as the activated experts are typically a small subset of the total experts. However, the small batch size results in low GPU utilization, as loading one expert can only process one or a few inputs, suffering from the transfer cost bottleneck.
Another branch of approach combats the sparsity of MoE layers by increasing the batch size, which allows the GPU to process multiple inputs simultaneously after loading the expert weights into the HBM. One representative work is MoE-Lightning~\cite{cao2025moe}, which processes a larger batch size once the expert weights are loaded into the HBM. However, as will be demonstrated in Section~\ref{sec:roofline_analysis}, the transfer cost of MoE-Lightning still dominates the performance under the realistic settings, as the CPU DRAM size may still limit the batch size.

\stitle{Offloading for KV cache.} KV cache and its corresponding attention computation are also critical components in MoE models, which are both memory consuming and memory-access intensive. Several works~\cite{sheng2023flexgen, kwon2023efficient} aim to leverage the external memory, e.g., DRAM or disk storage, to maintain the KV cache, thereby reducing the memory footprint of MoE models. When KV cache is required, the offloaded KV cache will be loaded back to the GPU HBM for attention computation. Despite saving the memory footprint, several works~\cite{agrawal2023sarathiefficientllminference} further propose CPU-based attention computation, as CPU has lower arithmetic intensity (Peak FLOPs / Peak Bandwidth) compared to GPU, which is suitable for the memory-bound attention computation. Again, MoE-Lightning~\cite{cao2025moe} adopts both techniques, i.e., offloading the KV cache to the CPU DRAM and performing the attention computation on the CPU, to reduce the HBM footprint and corresponding transfer cost, as well as to improve the CPU utilization.

\stitle{Our target scenario: CPU-GPU cooperated offloading.} We also notice that some recent works~\cite{ren2021zerooffloaddemocratizingbillionscalemodel} further explore the SSD/disk, memory pool, cloud storage, and other external memory resources to offload the MoE models. However, as will be investigated in Section~\ref{sec:roofline_analysis}, the transfer cost dominates the performance of MoE offloading even with the DRAM as external memory. Therefore, we focus on the CPU DRAM as the external memory and further explore the CPU computation.




\subsection{Speculative Decoding}
\label{sec:speculative_decoding}

To aid the insufficient GPU utilization in LLM inference, speculative decoding~\cite{xia2022speculative} has been proposed to improve the parallelism and reduce the memory access of KV cache in large language models.

As shown in Figure~\ref{fig:background}(c), speculative decoding consists of two main components: the draft model and the target model. In the draft phase, a lightweight draft model will be iteratively called to generate a sequence~\cite{leviathan2023fastinferencetransformersspeculative}/tree~\cite{chen2023acceleratinglargelanguagemodel} of draft tokens, which are then passed to the target model for verification. If the tokens are organized as a tree structure, an additional mask will be passed to the target model to represent the relationship between the draft tokens maintained as a sequence. In the verification phase, the original large language model, denoted as the target model, will verify the correctness of the draft tokens in one forward pass, which increases the parallelism and reduces the memory access of KV cache for verification of draft tokens.

Among various speculative decoding methods~\cite{leviathan2023fastinferencetransformersspeculative, chen2023acceleratinglargelanguagemodel,stern2018blockwiseparalleldecodingdeep,li2024eagle}, EAGLE~\cite{li2024eagle} is a representative work that employs lightweight draft models to generate draft tokens. EAGLE leverages the hidden state of the original target LLM as input and designs a one layer traditional transformer including an attention layer and an FFN as the draft model. In this paper, \name adopts the EAGLE framework to implement speculative decoding for demonstrating the effectiveness of speculative decoding in MoE offloading scenarios, while \name can be easily extended to other speculative decoding methods.

\section{Roofline Analysis of MoE Offloading}
\label{sec:roofline_analysis}

\subsection{Definition and Notations}
\label{sec:roofline_definition}
In the inference of MoE models, there are four major configurable factors: hardware, model, workload, and hyperparameters. Details of these factors are as follows:

\stitle{Hardware $\mathcal{H}$:} Considering the offloading scenario, in addition to the traditional GPU configurations, we need to consider the DRAM and CPU configurations. Specifically, we introduce $P_{\text{GPU}}$, $B_{\text{GPU}}$, $P_{\text{CPU}}$ and $B_{\text{CPU}}$ to denote the peak computational performance and memory bandwidth of the GPU and CPU, respectively. Moreover, we also introduce $B_{\text{CPU-GPU}}$ to denote the data transfer bandwidth between CPU and GPU. As the data transfer from GPU to CPU is negligible in MoE inference, $B_{\text{CPU-GPU}}$ will primarily represent the data transfer from CPU to GPU.

\stitle{Model $\mathcal{M}$:}
In the context of speculative decoding, the model specifies the target model $\mathcal{M}_{\text{target}}$ and the draft model $\mathcal{M}_{\text{draft}}$. As the target model dominates the overall processing time, we focus on it in our roofline analysis. We use $h$ and $h_\text{i}$ to denote the hidden dimension size and expert intermediate size. Regarding the MoE layer, we denote the number of experts as $n_{\text{expert}}$, the number of activated experts as $n_{\text{activate}}$, and the size of each expert as $e=h \times h_\text{i}$.
For the attention layer, the input and output hidden dimension sizes are also $h$, which will be split into multi heads. Moreover, we also use $g$ to represent the memory reduction factor of the KV cache, e.g., the group size in group-query attention~\cite{ainslie2023gqatraininggeneralizedmultiquery} combining $g$ query heads into one group of KV cache.
The model also includes the number of layers $l$ in the MoE model.

\stitle{Workload $\mathcal{W}$:}
A traditional parameter of the workload is the sequence length $s$. Moreover, the difficulty of tasks will also affect the performance under speculative decoding. Therefore, we use the acceptance rate function $a:\mathbb{R}\to[0,1]$ to denote the acceptance rate of the draft tokens, where $a(k)$ is the acceptance rate when $k$ draft tokens are generated. The acceptance rate can be obtained by profiling the prediction of the draft model.

\stitle{Hyperparameters $\mathcal{P}$:}
In traditional inference, the hyperparameters mainly include the batch size $b$ and the micro-batch size $m$. Speculative decoding further introduces the number of draft tokens $k$. Moreover, the hyperparameters also include the strategy $\mathcal{S}=\{ \mathcal{S}_\text{memory}, \mathcal{S}_\text{execution} \}$, which determines how the data is managed (e.g., cache policy) and how the computation is executed (e.g., scheduling).

\subsection{Cost Analysis}
Despite the visualization power of the roofline model, its main idea of rooflines is to compare the theoretical computation and memory access demand with the hardware's peak computational performance and memory bandwidth to determine the performance bottleneck of computation or memory access.
In this section, we analyze the cost of existing implementations of MoE layers and attention layers and proceed the visualization to the subsequent section.

\begin{table*}[htbp]
    \centering
    \caption{Hardware configurations and cost of executing operator implementations.}
    \scalebox{0.8}{
        \begin{tabular}{c|clllll}
            \hline
                                         & Hardware configuration                                     & $P_{\text{GPU}}$ (TFLOPS)                       & $B_{\text{GPU}}$ (GB/s)               & $B_{\text{CPU-GPU}}$ (GB/s)                               & $P_{\text{CPU}}$ (TFLOPS)          & $B_{\text{CPU}}$ (GB/s)               \\ \hline
            \multirow{2}[0]{*}{Hardware} & \multicolumn{1}{c}{A30 + INTEL Gold 6426Y}                 & 165                                             & 933                                   & 25                                                        & 2.43                               & 357                                   \\
                                         & \multicolumn{1}{c}{4090D + INTEL Gold 5418Y}               & 83                                              & 1008                                  & 23                                                        & 1.45                               & 197                                   \\ \hline
                                         & Operator implementation                                    & GPU Operations                                  & \multicolumn{1}{l}{GPU Memory access} & \multicolumn{1}{l}{Memory transfer}                       & \multicolumn{1}{l}{CPU Operations} & \multicolumn{1}{l}{CPU Memory access} \\ \hline
            \multirow{4}[0]{*}{Software} & MoE (large batch)~\cite{cao2025moe}                        & $3\times e \times n_{\text{activate}} \times b$ & $3\times n_{\text{expert}} \times e$  & $3\times n_{\text{expert}} \times e$                      & 0                                  & 0                                     \\
                                         & MoE (batch=1)~\cite{yu2025fmoefinegrainedexpertoffloading} & $3\times e \times n_{\text{activate}}$          & $3\times n_\text{activate} \times e$  & $3\times n_\text{activate} \times e \times r_\text{miss}$ & 0                                  & 0                                     \\
                                         & attention (in CPU)~\cite{cao2025moe}                       & 0                                               & 0                                     & 0                                                         & $2\times b\times s\times h$        & $2\times b\times s\times h/g$         \\
                                         & attention (to GPU)~\cite{sheng2023flexgen}                 & $2\times b\times s\times h$                     & $2\times b\times s\times h/g$         & $2\times b\times s\times h/g$                             & 0                                  & 0                                     \\ \hline
        \end{tabular}%
    }
    \label{tab:hardware_and_roofline}%
\end{table*}%

Table~\ref{tab:hardware_and_roofline} summarizes the hardware configurations and the cost of executing the operator implementations. We discuss them in detail in the following sections.

\stitle{MoE layer:}
The computation of an expert in MoE layer, i.e., feed-forward network (FFN), mainly involves three matrix multiplications as follows:
\begin{align*}
    \mathrm{FFN}(x) = \mathbf{W}_{\text{down}} \Big( \sigma \big( \mathbf{W}_{\text{gate}} x \big) \odot \big( \mathbf{W}_{\text{up}} x \big) \Big),
\end{align*}
where $\mathbf{W}_\text{up},\mathbf{W}_\text{gate} \in \mathbb{R}^{h \times h_\text{i}}$ and $\mathbf{W}_\text{down} \in \mathbb{R}^{h_\text{i} \times h}$.
The computation cost of three matrix multiplications can be summarized $3 \times b \times e$, where $e=h\times h_\text{i}$.
Regarding the memory access, we follow~\cite{sheng2023flexgen} by using input matrix size to represent the memory access cost, which is $e+b\times h$, as the size of the hidden states is negligible compared to the expert weights in decoding stage (e.g., 0.006 GB hidden states while 2.6 GB matrix size in our case), we use $e$ to denote the memory access of an expert in Table~\ref{tab:hardware_and_roofline}.

Regarding the computation of a batch of input tokens, as each token activates $n_{\text{activate}}$ experts, the total computation cost of the MoE layer can be easily derived as $n_{\text{activate}} \times b \times e$. In contrast, the memory access cost is sensitive to the batch size $b$ and its corresponding implementation. For large batch size, we assume all experts are activated, and the memory access cost and transfer cost are $n_{\text{expert}} \times e$. Regarding the small batch size, only the activated experts are accessed, and the memory access cost is $n_{\text{activate}} \times e$, while the caching and prediction mechanism of experts can further reduce the memory transfer cost by a factor of $r_{\text{miss}}$, which is the cache miss rate. Obviously, regarding MoE layer in throughput-oriented task, we have:
\begin{center}
    \framebox{
        \begin{minipage}{0.9\linewidth}
            \centering
            \emph{Large batch is better when $b\geq \frac{n_{\text{expert}}}{n_{\text{activate}\times r_{\text{miss}}}}$,}
        \end{minipage}
    }
\end{center}
which can be easily achieved in practice. Therefore, we focus on the large batch size in the following analysis.

\stitle{Attention layer:}
The cost of the attention layer is dominated by the computation of the attention scores and the weighted sum of values. In the decoding stage, for each request, only one token will be processed at a time, leading to GEMV (General Matrix-Vector multiplication) operations, which have the same computation and memory access cost, $s\times h$. Given $b$ requests, the total computation cost and memory access of the attention layer is $2 \times b \times s \times h$.
To facilitate the discussion, we omit the cost of query matrix, as it is negligible compared to the key and value matrices in decoding stage. We also notice multi-query attention~\cite{shazeer2019fasttransformerdecodingwritehead}, grouped query attention~\cite{ainslie2023gqatraininggeneralizedmultiquery}, cross layer attention~\cite{dai2019transformerxlattentivelanguagemodels} and multi-head latent attention~\cite{jaegle2021perceivergeneralperceptioniterative} can further optimize the memory access and consumption of the attention layer, we introduce the $g$ (group size in group-query attention) to represent the size reduction factor of the attention layer compared to multi-head attention.  In morden hardware, $B_{\text{CPU}}$ typically larges than $B_{\text{CPU-GPU}}$, thus we can conclude that
\begin{center}
    \framebox{
        \begin{minipage}{0.9\linewidth}
            \centering
            \emph{CPU-based attention is better than transferring to GPU.}
        \end{minipage}
    }
\end{center}

\subsection{Roofline Visualization}\label{sec:roofline_visual}

To give a intuitive understanding of bottlenecks in existing solutions, we conduct the roofline analysis of the MoE layer and attention layer in Mixtral 8x7B model on two hardware configurations: A30 and 4090D, each cofigured with 250 and 190 GB CPU memory, respectively.
The model consumes 87GB of memory in FP16 precision, leaving approximately 160 GB and 100 GB of CPU memory for KV cache, respectively.
Usually, $b$ is limited by the CPU memory size, because all requests' KV cache needs to be placed in CPU memory.
The workload we used to analyze is APPS, which is a popular coding dataset with an average input length of 566, as shown in Table~\ref{tab: dataset_table}. The output length is set to 1024, which is normal in code generation tasks.
Assuming the KV cache adopts FP16 precision, the global batch sizes of A30 and 4090D environments are 824 and 515, respectively.


\begin{figure}[h]
    \centering
    \begin{subfigure}{0.49\linewidth}
        \includegraphics[width=\linewidth]{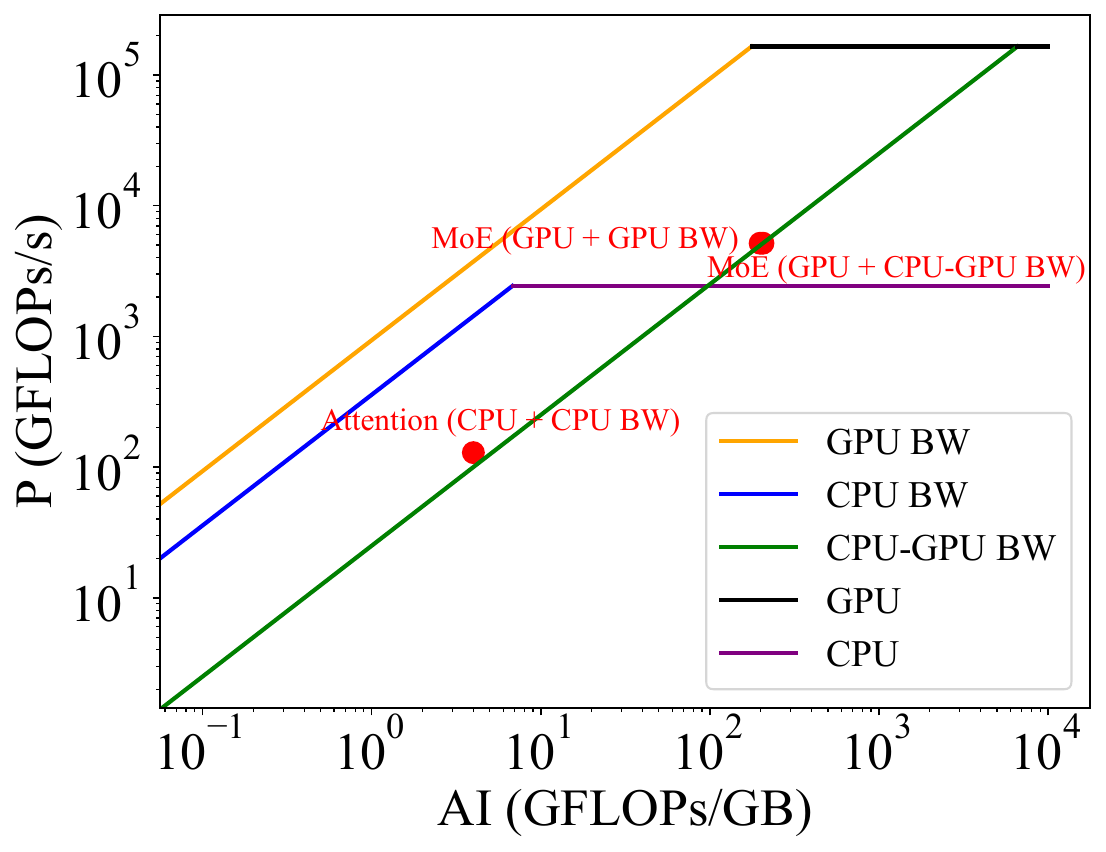}
        \caption{Roofline model for A30}
    \end{subfigure}
    \begin{subfigure}{0.49\linewidth}
        \includegraphics[width=\linewidth]{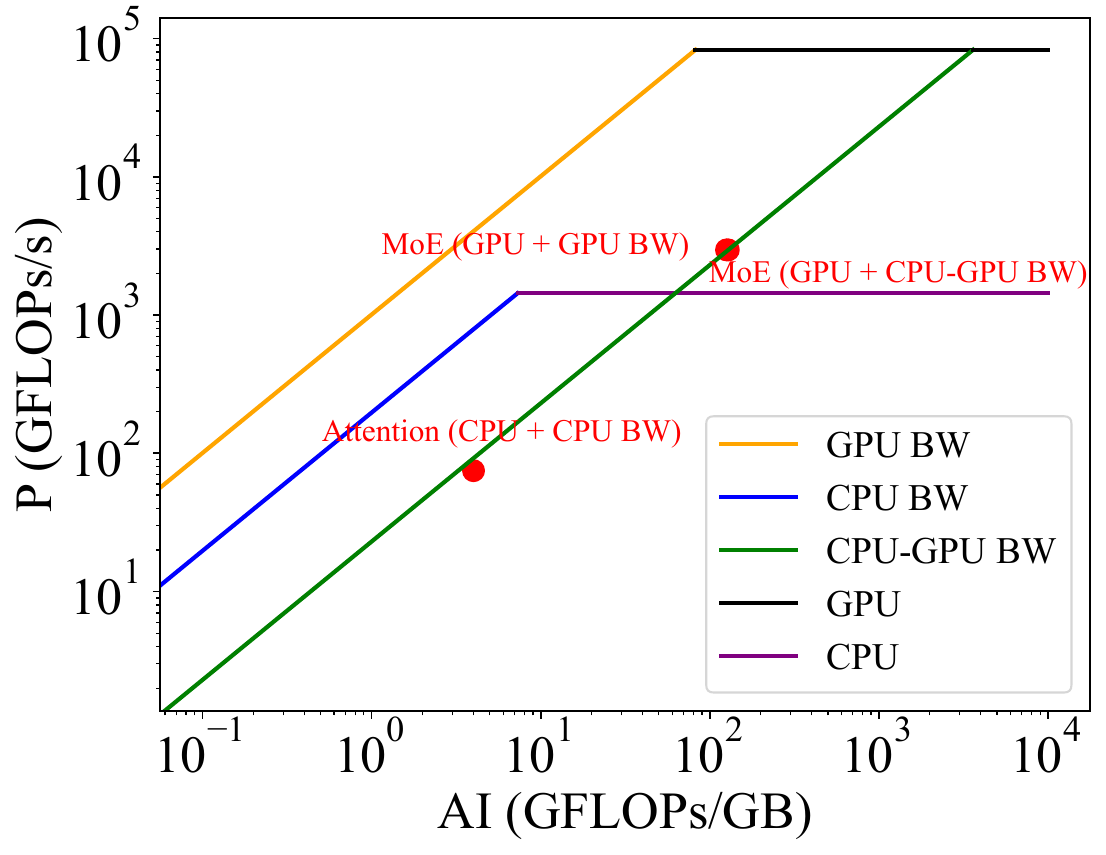}
        \caption{Roofline model for 4090D}
    \end{subfigure}
    \caption{Hierarchical Roofline Models for Mixtral 8x7B in large batch size on A30 and 4090D instances.}
    \label{fig:roofline}
\end{figure}

Figure~\ref{fig:roofline} presents the hierarchical Roofline models. In the model, the horizontal lines represent the compute roofs defined by the peak performance of the GPU and CPU, while the memory roofs are determined by the GPU memory bandwidth, CPU memory bandwidth, and the CPU-to-GPU memory transfer bandwidth.
For the two major computational kernels in the model—MoE and Attention layers—their performance points are plotted based on the corresponding operational intensities and arithmetic powers.
Notably, the MoE layer computation is performed solely on the GPU, with memory access involving only GPU memory and CPU-to-GPU transfers, resulting in two performance points. Similarly, the Attention layer is offloaded to the CPU, so only CPU compute and memory access are considered, corresponding to a single performance point.

Specifically, from the Roofline models, we observe that despite minor differences across devices, the performance bottlenecks for MoE and Attention computations remain consistent. For the MoE layer on A30 GPU, the GPU compute and memory performance point falls within the compute-bound region, but the arithmetic utilization is only 3.13\%. In contrast, the GPU compute and CPU-GPU transfer point is situated in the memory-bound region, with the transfer bandwidth nearly fully utilized.

In summary, \emph{the MoE layer is primarily limited by the CPU-GPU memory transfer for expert weights, while the Attention layer is limited by CPU memory access for KV cache.} Fortunately, the increased degree of parallelism and reduced memory access of KV cache in speculative decoding well mitigates these bottlenecks.

\section{Overview}
\label{sec:overview}

\subsection{System Architecture}
\label{sec:system_architecture}

\name has four main components: the target model execution engine, the draft model execution engine, the hyperparameter optimizer, and the memory manager.

\stitle{Target Model Execution Engine.}
The main component of the target model execution engine is similar to the existing MoE offloading systems, such as MoE-Lightning~\cite{cao2025moe}, which involve a GPU kernel for the MoE layer and a CPU kernel for the attention layer.

\stitle{Draft Model Execution Engine.}
Due to the limited GPU HBM capacity and the large KV cache size of the draft model in throughput-oriented scenarios, the draft model execution engine still needs to offload some of the KV cache to CPU DRAM. Details of the draft model execution engine described in Section~\ref{sec:memory_conscious_draft_model_execution}.

\begin{figure}
    \centering
    \includegraphics[width=0.95\linewidth]{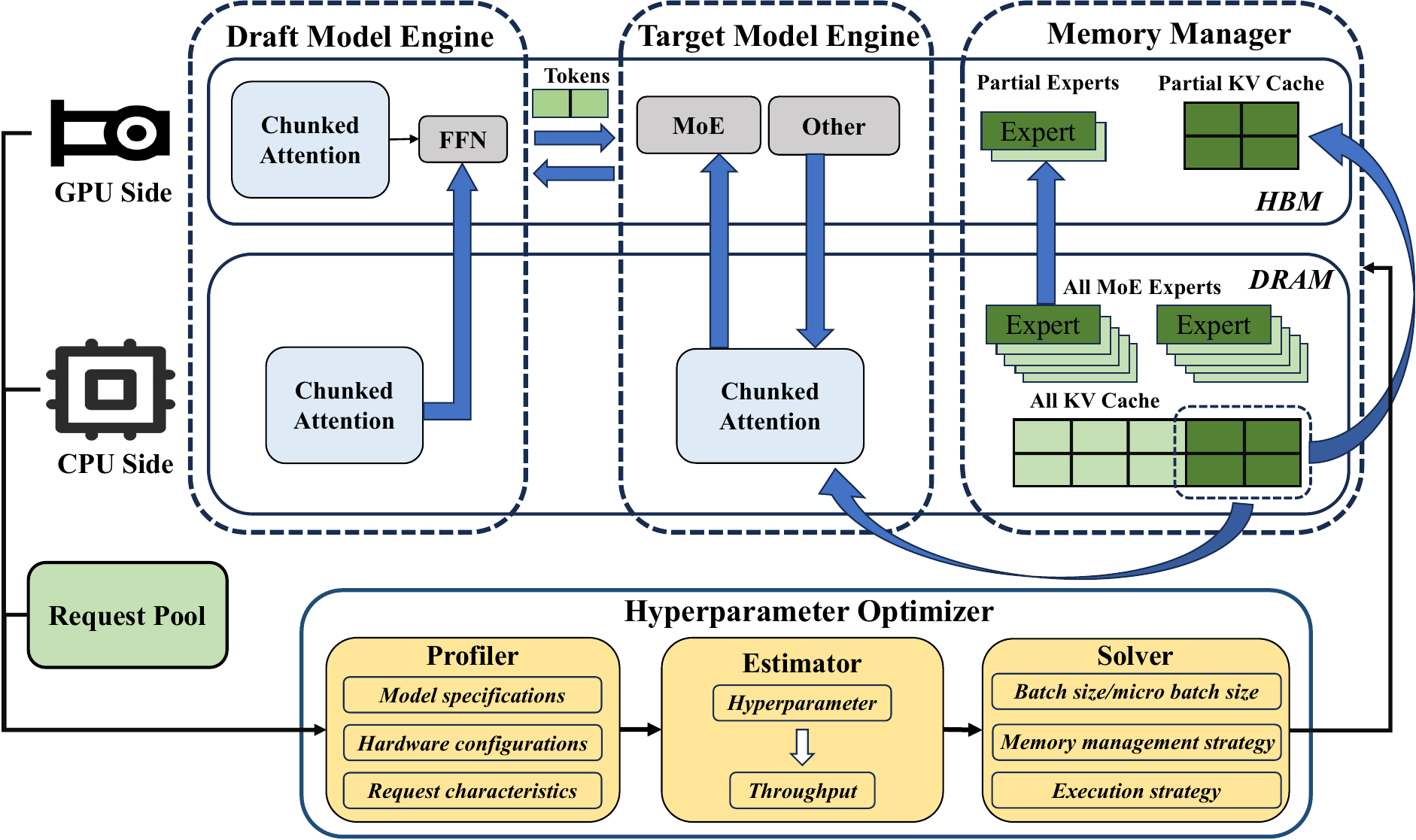}
    \vspace{-0.1in}
    \caption{System Architecture of \name.}
    \label{fig:system_architecture}
    \vspace{-0.2in}
\end{figure}

\stitle{Hyperparameter Optimizer.}
The hyperparameter optimizer is responsible for tuning the offloading and speculative decoding hyperparameters in \name.
As shown in Figure~\ref{fig:system_architecture}, we first collect the hardware configuration and model specifications, then conduct micro-benchmarking for the profiler to collect the performance of different hyperparameters. The estimator will estimate the performance of \name based on the collected data, and the solver will solve the convex optimization problem to find the optimal hyperparameters. Then the hyperparameter optimizer will generate the execution plan for \name.

\stitle{Memory Manager.}
This component is responsible for managing the memory resources within the GPU and CPU, which will be guided by the hyperparameter optimizer to determine the memory allocation strategy for the target and draft models, including the KV cache size and the model parameters placement.

\subsection{Execution Workflow}


As the prefill stage is naturally compute bound, which has been well studied in existing MoE offloading systems, we do not focus on optimizing it and discuss it briefly here.
We chunk global batch into multiple micro-batches. Load one layer's expert parameters and the former layer's output hidden states to GPU, iterate the micro-batches and execute the layer's computation. After the computation of one micro-batch, we offload both the KV Cache and hidden states to CPU DRAM.
With regard to the draft model, we put all parameters on GPU HBM, and conduct GPU-based prefill.

We now focus on the decoding stage.
The target model is executed first to verify the correctness of the draft tokens, followed by the draft model to update the KV cache according to the target verification results, and generate new draft tokens. As the data transfer of activations and generated QKV is negligible, we omit them in the following discussion.

\begin{figure}[t]
    \centering
    \includegraphics[width=\linewidth]{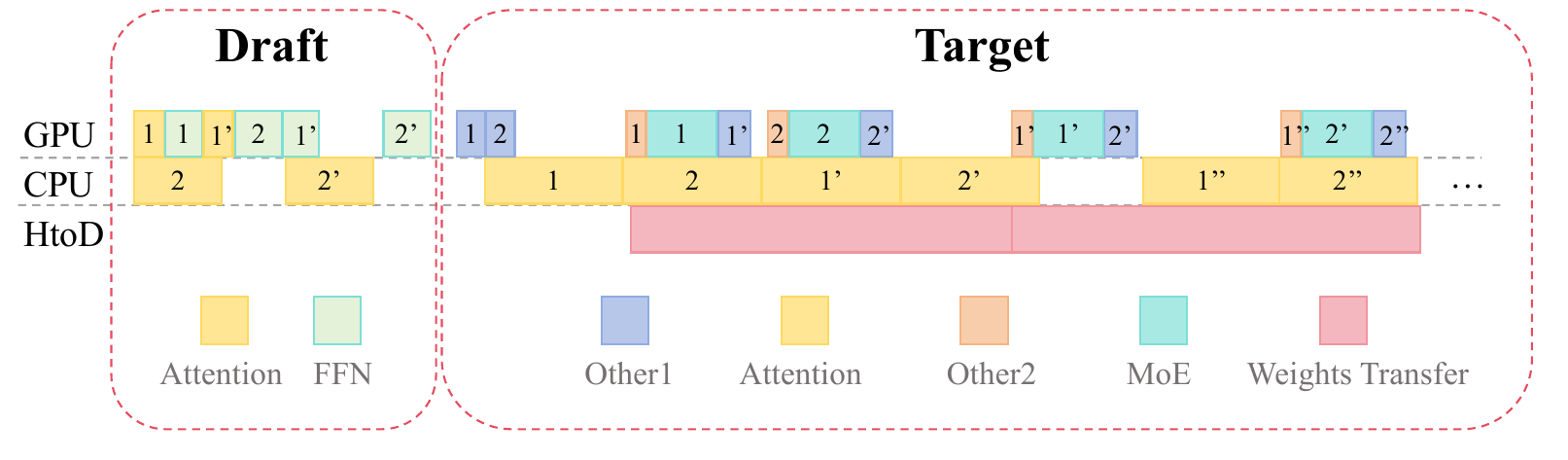}
    \vspace{-0.1in}
    \caption{\name execution pipeline.}
    \label{fig:pipeline}
    \vspace{-0.2in}
\end{figure}

\stitle{Draft model execution.}
Since GPU memory cannot accommodate the entire draft KV cache, we split the batch into two parts: the CPU Part and the GPU Part. As illustrated in the left part of Figure~\ref{fig:pipeline}, the numbers 1 and 2 represent the GPU Part and CPU Part, respectively, while 1$^{\prime}$ and 2$^{\prime}$ denote the second forward pass. The attention computations of the GPU Part and CPU Part are performed parallelly, while the both FFN computations are executed on the GPU as GPU still retains unused computational resources.


\stitle{Target model execution.} As shown in the right part of Figure~\ref{fig:pipeline}, we perform CPU Attention computation, and carefully arrange CPU, GPU and I/O pipeline which is similar to MoE-Lightning~\cite{cao2025moe}.
During inference, only the experts need to be transferred to the GPU for computation. As illustrated in the Figure~\ref{fig:pipeline}, we split the batch size into two micro-batches. The numbers 1 and 2 denote micro-batch IDs, and 1$^{\prime}$ and 2$^{\prime}$ indicate computation in the next layer, and so on. Each micro-batch sequentially executes GPU Other1, CPU Attention, GPU Other2, and GPU MoE computation. This design enables overlapping of CPU and GPU computations between the two micro-batches. While computing layer \textit{i}, the expert weights for layer \textit{i+1} are transferred concurrently.

\section{Unique Components in \name}
\label{sec:unique_components}
As there are various works on MoE offloading, we describe the unique features in the design and implementation of \name in this section.


\subsection{Chunked Attention in Target Model Verification}
\label{sec:chunked_attention}

The target model verification is to verify the correctness of a chunk of draft tokens generated by the draft model, which varies from the original target model inference that only autoregressively generates one token at a time. Though this process improves the degree of parallelism and combines the memory access of KV cache, it also introduces new challenges for implementation, especially in the attention operator. Notice that the MoE layers and other layers in the target model operate solely on the current or to-be-verified tokens, and thus can be easily extended from standard decoding implementations. In contrast, the KV cache access for the chunk of draft tokens needs to be combined, which needs to be carefully designed to avoid the memory access bottleneck. To facilitate the understanding, we will use a single request to illustrate the design of the chunked attention operator, while extending it to batched requests can be seamlessly done by parallelizing the computation across multiple requests.


\begin{figure}[htbp]
    \centering
    \includegraphics[width=0.8\linewidth]{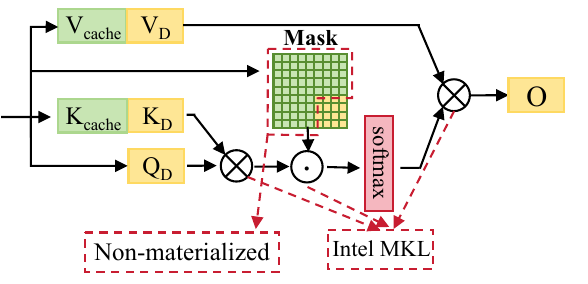}
    \vspace{-0.12in}
    \caption{Illustration of chunked attention operator.}
    \label{fig:chunked_attention}
    \vspace{-0.2in}
\end{figure}
\stitle{Chunked Attention.}
As illustrated in figure~\ref{fig:chunked_attention}, the chunked attention operator involves two steps.
The first step is to compute the attention scores between the query of chunked draft tokens $Q\in\mathbb{R}^{n\times d}$ and the key of the all tokens including the draft tokens and previous tokens $K\in\mathbb{R}^{(l+n)\times d}$, where $n$ is the number of draft tokens, $l$ is the number of previous tokens, and $d$ is the hidden dimension. The second step is to compute the attention output by multiplying the attention scores with the value of all tokens $V\in\mathbb{R}^{(l+n)\times d}$ under the attention mask $M\in\{-\infty,1\}^{n\times (l+n)}$, where $M_{ij}=0$ indicates that the $i$-th draft token can attend to the $j$-th token, and $M_{ij}=-\infty$ indicates that the $i$-th draft token cannot attend to the $j$-th token.


\stitle{Naive solution for DRAM-maintained chunked attention.}
A straightforward approach is to directly reuse existing attention operators on either GPU or CPU, yet each option incurs significant inefficiencies. Specifically, employing the GPU chunked attention operator~\cite{dao2022flashattention,wang2025flashforgeultraefficientprefixawareattention} requires repeatedly transferring the KV cache from CPU DRAM to GPU HBM, which introduces substantial transfer overhead. Alternatively, extending the CPU decoding attention operator~\cite{cao2025moe} demands repeating the operation for every draft tokens, leading to excessive memory access overhead and redundant computation. Similarly, extending the CPU prefill attention operator~\cite{pytorch} like Pytorch dose, results in duplicated computation for tokens that are already stored in the KV cache. Moreover, existing designs typically store the entire attention mask matrix for every decoding round. Since a large portion of its elements are constant (e.g., fixed to 1), this practice causes considerable memory waste.


\stitle{Our chunked CPU attention operator.}
To address the limitations of existing attention implementations in offloading scenarios, we propose a dedicated chunked CPU attention operator. Our operator is designed to efficiently handle multiple draft tokens while minimizing memory consumption and avoiding duplicated computation.

Different from the existing GPU-based attention implementations, our chunked CPU attention operator needs to utilize the CPU processing power and CPU cache abilities, which varies from the SIMT architecture and manually controlled shared memory in GPU. Specifically, we leverage the Intel CPU MKL library~\cite{intel_mkl} for efficient matrix operations to fully leverage CPU SIMD and MIMD capabilities for high performance. Moreover, since the parts of the mask matrix unrelated to Draft tokens (highlighted by the red dashed region in Figure~\ref{fig:chunked_attention}) are fixed to 1, we avoid materializing them and only store the portions relevant to Draft tokens, which substantially reduces memory requirements.

\subsection{Memory-Conscious Draft Model Execution}
\label{sec:memory_conscious_draft_model_execution}

Compared with the target model, the draft model is relatively small, but it still has a non-negligible memory footprint for KV cache, especially in our throughput-oriented scenarios with large batch size and long sequence length.
For instance, the eagle model for mixtral 8x7B only consumes 1.6 GB memory for parameters, while in 250 GB CPU memory setting, the target model's KV cache consumes 142 GB and the draft model's KV cache consumes 17.75 GB.
The GPU memory must accommodate a 5.25 GB of expert cache, a 1.63 GB of draft model, and 1 GB of target model activations. A 24 GB GPU is already insufficient to accommodate the draft KV cache under such conditions.
In contrast, under 4090D environment with 190 GB DRAM, the draft model KV cache is nearly 10 GB and all draft KV cache can be stored in GPU.


In addition, the execution of the draft model also varies from the target model, as the draft model needs to generate multiple draft tokens in each iteration, while the target model only needs to forward once to verify the correctness of the draft tokens.
This repeated computation pattern actually increases the memory access and computation cost by multiple times.
Considering the roofline analysis, under the same execution logic, loading or caching one piece of data (despite KV cache, hidden states, or parameters) on GPU HBM for draft model can achieve multiple times higher arithmetic intensity compared with loading the same piece of data for target model.

In summary, \emph{the data of the draft model should prioritize to utilize the GPU HBM, while the case of overflowing the GPU HBM should be carefully considered for the draft model execution.}

As analyzed in the roofline model and in the execution of the target model, the execution of the MoE layers (FFN in EAGLE~\cite{li2024eagle}) and other layers should be prioritized to be executed on GPU. Given the small size of the draft model, we by default put all the draft model parameters on GPU HBM. Subsequently, we focus on the KV cache as well as its corresponding attention operator in draft model.

\stitle{CPU/GPU separation for Draft Model Execution.}
As aforementioned, we aim to prioritize the draft model's KV cache to be placed on GPU HBM. We first consider how to partition the KV cache as well as the corresponding attention operator into two parts, one on CPU DRAM and the other on GPU HBM. As indicated in~\cite{wang2025flashforgeultraefficientprefixawareattention, dao2022flashattention}, the K cache or V cache is a 4D tensor, (formally, $K,V\in\mathbb{R}^{b\times h\times s\times d}$, where $b$ is the batch size, $h$ is the number of heads, $s$ is the sequence length, and $d$ is the hidden dimension), which can be partitioned in batch, head, or sequence dimension. However, partitioning in the head dimension need synchronization to gather the hidden states sourcing from different heads, while partitioning in the sequence dimension requires additional computation to combine the partial attention scores sourcing from different sequence parts. Therefore, we choose to partition the KV cache in the batch dimension, which can be easily implemented by slicing the KV cache into multiple chunks, each of which is a 4D tensor.

\begin{figure}[t]
    \centering
    \includegraphics[width=\linewidth]{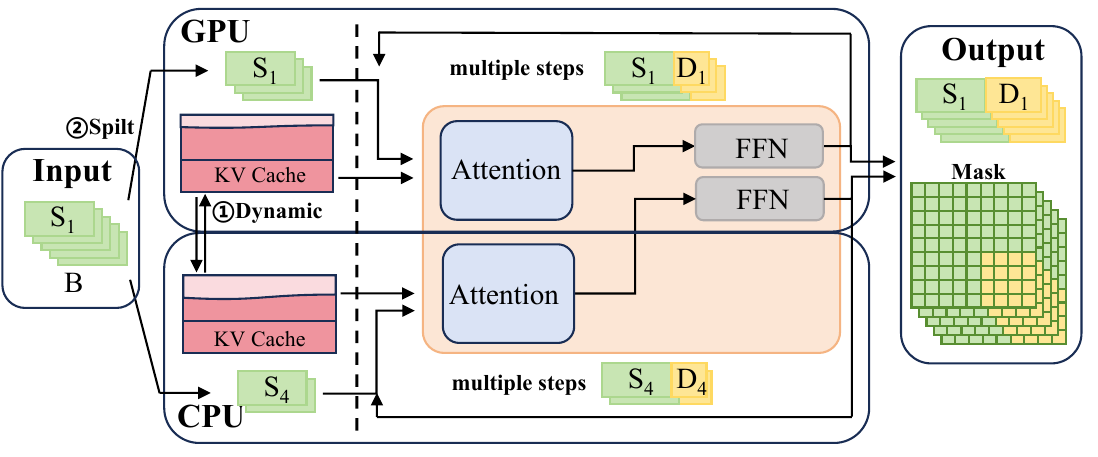}
    \vspace{-0.2in}
    \caption{Memory-conscious draft model execution.}
    \vspace{-0.2in}
    \label{fig:memory_conscious}
\end{figure}

As illustrated in figure~\ref{fig:memory_conscious}, to avoid the synchronization overhead between two parts of requests, \name separates the execution of the whole draft model instead of only the attention operator. Specifically, we have follwing two distinct kernels executed parallelly:
\begin{itemize}
    \item \textbf{GPU-only draft model execution:} This kernel executes the draft model on directly on GPU, where the corresponding KV cache is fully placed on GPU memory.
    \item \textbf{CPU/GPU cooperation draft model execution:} This kernel executes the attention operation on CPU and transfers the hidden states to GPU for subsequent FFN layer computation, the computation is repeated until all the draft tokens are generated.
          Notice that the computation of the attention operator is also performed on CPU, which is similar to the target model verification.
\end{itemize}

We also develop a dynamic separate ratio mechanism to minimize the draft execution time. During the initial phase of a generation task, more requests can be scheduled for execution on the GPU. As generation progresses and the sequence length increases, the draft KV cache eventually becomes insufficient. At this point, some requests are offloaded to the CPU. Conversely, when some requests are completed, the system dynamically loads the draft KV cache of other requests back into the GPU for execution.

\subsection{Hyperparameter Optimizer}
\label{sec:hyperparameter_optimizer}

As indicated in both offloading systems~\cite{cao2025moe,sheng2023flexgen} and speculative decoding algorithms~\cite{li2024eagle}, the hyperparameters, including batch size, micro-batch size, the number of draft tokens generated in each iteration, and etc., are critical for the performance of the system. In \name, we design a hyperparameter optimizer to determine the optimal hyperparameters for the execution of the target and draft models.

\stitle{Optimization Problem.}
Our target is to maximize the throughput of the system under the given hardware configurations $\mathcal{H}$, model specifications $\mathcal{M}$, and request characteristics $\mathcal{R}$. We aim to find the optimal hyperparameter $\mathcal{P}=(b, m, k, \mathcal{S}$, including the batch size $b$, micro-batch size $m$, the number of draft tokens $k$, the strategies $\mathcal{S}$ including the memory management policy $\mathcal{S}_\text{memory}$, and the execution strategies $\mathcal{S}_\text{execution}$, to maximize the throughput $T$ of the system:
\begin{align}
    \text{Given } \mathcal{H}, \mathcal{M}, \mathcal{R}, \text{ find } \mathcal{P}^* = \mathop{\arg\max}_{\mathcal{P}=(b, m, k, \mathcal{S})} T(\mathcal{H}, \mathcal{M}, \mathcal{R}, \mathcal{P}).
\end{align}

There are following chanllenges in solving the optimization problem:
1) It is impractical to validate all the possible combinations of hyperparameters.
2) The estimation of the throughput found by theoretical models is inaccurate~\cite{zhou2024survey}. 

In contrast to the traditional theoretical-based approaches~\cite{cao2025moe}, we adopt a hybrid approach that combines the convex optimization and profiling-based methods to solve the optimization problem. Specifically, we first use the theoretical models to reduce the search space of hyperparameters, and then use the profiling-based methods to find the optimal hyperparameters in the reduced search space.

\stitle{Convex Optimization.}
Actually, several hyperparameters, such as the batch size, memory management policy and execution strategy, can be determined by the theoretical models in roofline analysis and our aforementioned design philosophy.
1) Larger batch size is preferred to increase the operational intensity of the computation. So we set the batch size $b$ to the maximum value that can fit in the CPU memory.
2) Maximizing the GPU cache for draft model execution. For further available GPU HBM, we choose to put the KV cache of the target model on GPU HBM, so we can get $\mathcal{S}_\text{memory}$ when GPU memory size is provided.

\stitle{Profiling-based Estimator.}
Before deploying the given model specification $\mathcal{M}$ on the target hardware configuration $\mathcal{H}$, we first conduct a micro-benchmarking to profile the performance of the target model and draft model on the given hardware, the profiled operators including GPU MoE, GPU Attention, CPU Chunked Attention and HtoD Transfer. Then we use linear model to fit the profiled performance data to form performance models. We conduct offline experiments with diverse datasets to evaluate the draft model's capability, ultimately establishing a mapping from the number of draft tokens to the number of accepted tokens.

We developed a simulator that, given hyperparameters $\mathcal{P}$, models the GPU, CPU, and I/O pipeline, and constructs a DAG graph. Each node in the graph represents an event, such as GPU MoE computation. There are many dependencies between nodes, for example, MoE operation must wait for CPU Chunked Attention results generated and expert weights are transferred to GPU. We use topological sorting to traverse the DAG, determining the final execution time.

Regarding other overheads, such as the verification of draft tokens after target model execution, they can be estimated using profiling methods. These overheads are positively correlated with the batch size and the draft tokens number $k$.

The number of tokens generated in a single iteration is determined by consulting the draft token number to accepted token mapping. Thus we get the throughput $T$ of the system give the hyperparameters $\mathcal{P}$.



\stitle{Problem Solving.}
We adopt the convex optimization to predecide some hyperparameters, including batch size $b$, micro-batch size $m$, memory management policy $\mathcal{S}_\text{memory}$, and execution strategy $\mathcal{S}_\text{execution}$.
The hyperparameters $\mathcal{P}$ only has one variable $k$ to be determined, and its range is limited. So we vary $k$ to find the optimal execution plan for the target and draft models. $k$ along with the predecided hyperparameters are given to the estimator to get estimated throughput. Then we choose the largest throughput and its corresponding hyperparameters as the final solution.





\section{Evaluation}
\label{sec:evaluation}

\subsection{Experimental Setup}
\stitle{Implementation.} We implement \name on top of SGLang~\cite{zheng2024sglangefficientexecutionstructured} and adopt several designs (e.g., FFN, expert cache) from MoE-Lightning~\cite{cao2025moe}, with 20,000+ additional LoC of Python/C++/CUDA.
For concurrent data transfers and computation, we adopt separated CUDA Streams for GPU computation, expert weight loading, activation loading and activation offloading, and use CUDA Events to synchronize different streams.
To efficiently manage the memory resources and accelerate the data transfer, we employ pin memory
and dynamic memory allocation for demanded tensors, which reduces the memory copy overhead and avoids the memory fragmentation problem.

\stitle{Hardware $\mathcal{H}$.}
We use same hardware environments shown in Table~\ref{tab:hardware_and_roofline}. Due to the cloud service limitations, we configure the CPU memory as 250 for environment 1 (abbr. A30) and 190 for environment 2 (abbr. 4090D).


\stitle{Models $\mathcal{M}$.}
We employed the popular Mixtral-8x7B MoE model~\cite{jiang2024mixtral} to evaluate \name, utilizing its EAGLE~\cite{li2024eagle} model as the draft model. At FP16 precision, the Mixtral-8x7B model occupies 87 GB of memory, with its experts consuming 84 GB and the non-expert components accounting for the remaining 3 GB. Its corresponding EAGLE model occupies less than 2 GB of memory.

\stitle{Workload $\mathcal{W}$.}
Given that different tasks exhibit distinct input characteristics and varying draft token acceptance rates, we employ two popular but different tasks in our evaluation, as detailed in Table~\ref{tab: dataset_table} ($s$ only includes the input sequence length).
The CNNDailyMail dataset is widely used for tasks such as text summarization, reading comprehension, and question answering. It consists of news articles sourced from the Cable News Network (CNN) and the Daily Mail website.
The APPS (Automated Programming Progress Standard) dataset is a large-scale, challenging dataset designed to evaluate the capabilities of code generation models. It comprises 10,000 programming problems.
In addition to varying input characteristics, we also control the output length of the generated tokens, which is a common practice in API-based inference systems~\cite{brown2020languagemodelsfewshotlearners} and evaluation benchmarks~\cite{liang2023holisticevaluationlanguagemodels}.

\begin{table}[]
    \centering
    \caption{Datasets, where $\mu_s$ and $\sigma_s$ denote the average and standard deviation of the sequence lengths, respectively.}
    \resizebox{\columnwidth}{!}{%
        \begin{tabular}{cccc}
            \hline
            Dataset                           & Task             & $\mu_s$ & $\sigma_s$ \\ \hline
            APPS~\cite{hendrycksapps2021}     & Coding           & 566.17  & 271.80     \\
            CNN/DailyMail~\cite{cnndailymail} & Summarization/QA & 1005.56 & 491.66     \\ \hline
        \end{tabular}
    }
    \label{tab: dataset_table}
\end{table}







\subsection{End-to-End Comparisons}

\begin{figure*}[t]
    \centering
    \includegraphics[width=0.8\linewidth]{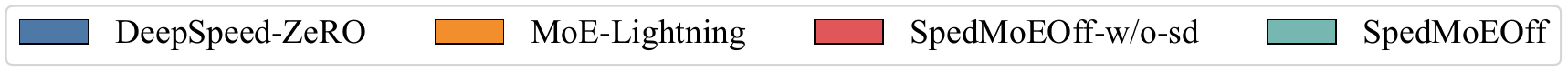}
    \vfill
    \begin{subfigure}[t]{0.24\linewidth}
        \centering
        \includegraphics[width=\linewidth]{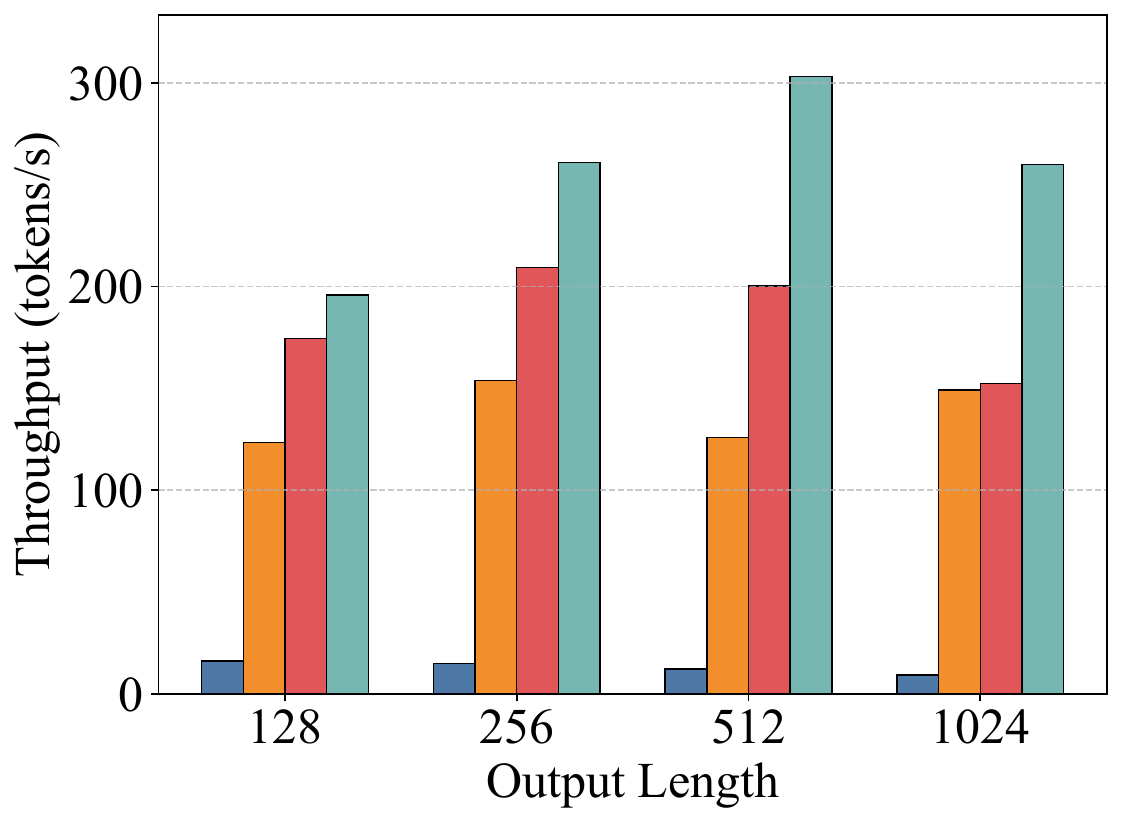}
        \caption{A30 APPS dataset}
    \end{subfigure}
    \begin{subfigure}[t]{0.24\linewidth}
        \centering
        \includegraphics[width=\linewidth]{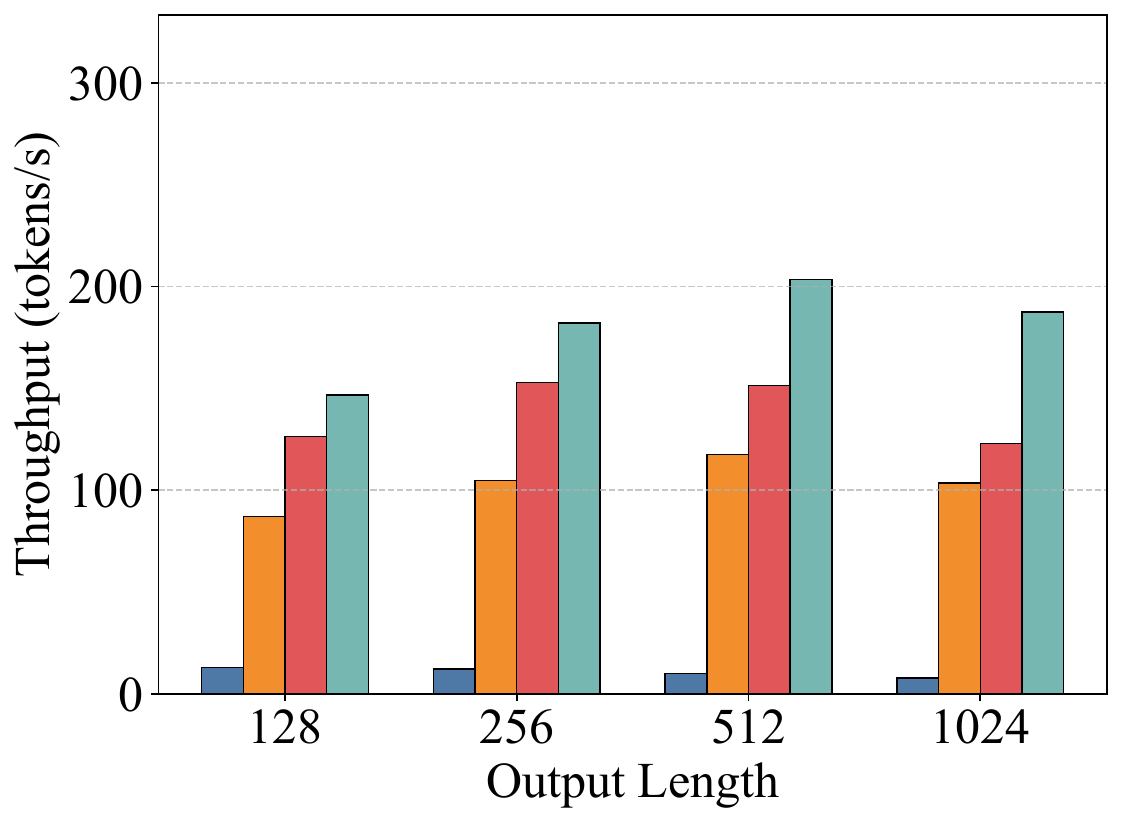}
        \caption{A30 CNN/DailyMail dataset}
    \end{subfigure}
    \begin{subfigure}[t]{0.24\linewidth}
        \centering
        \includegraphics[width=\linewidth]{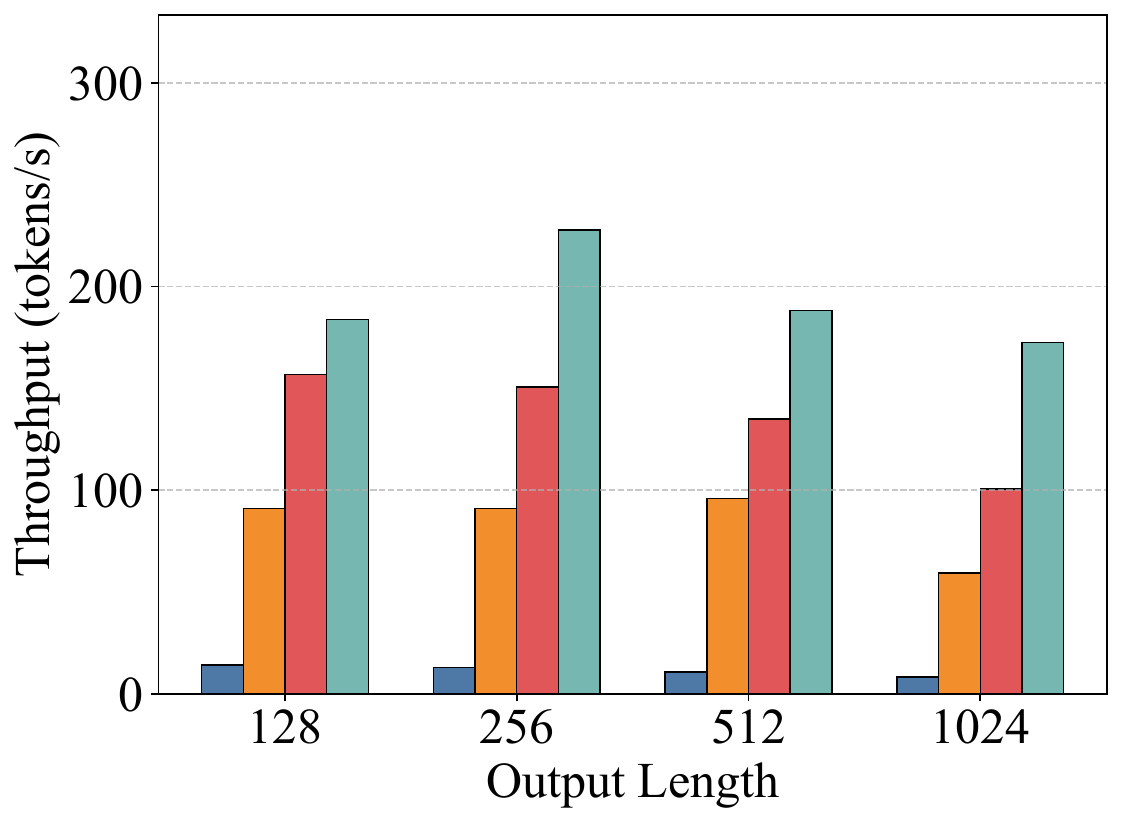}
        \caption{4090D APPS dataset}
    \end{subfigure}
    \begin{subfigure}[t]{0.24\linewidth}
        \centering
        \includegraphics[width=\linewidth]{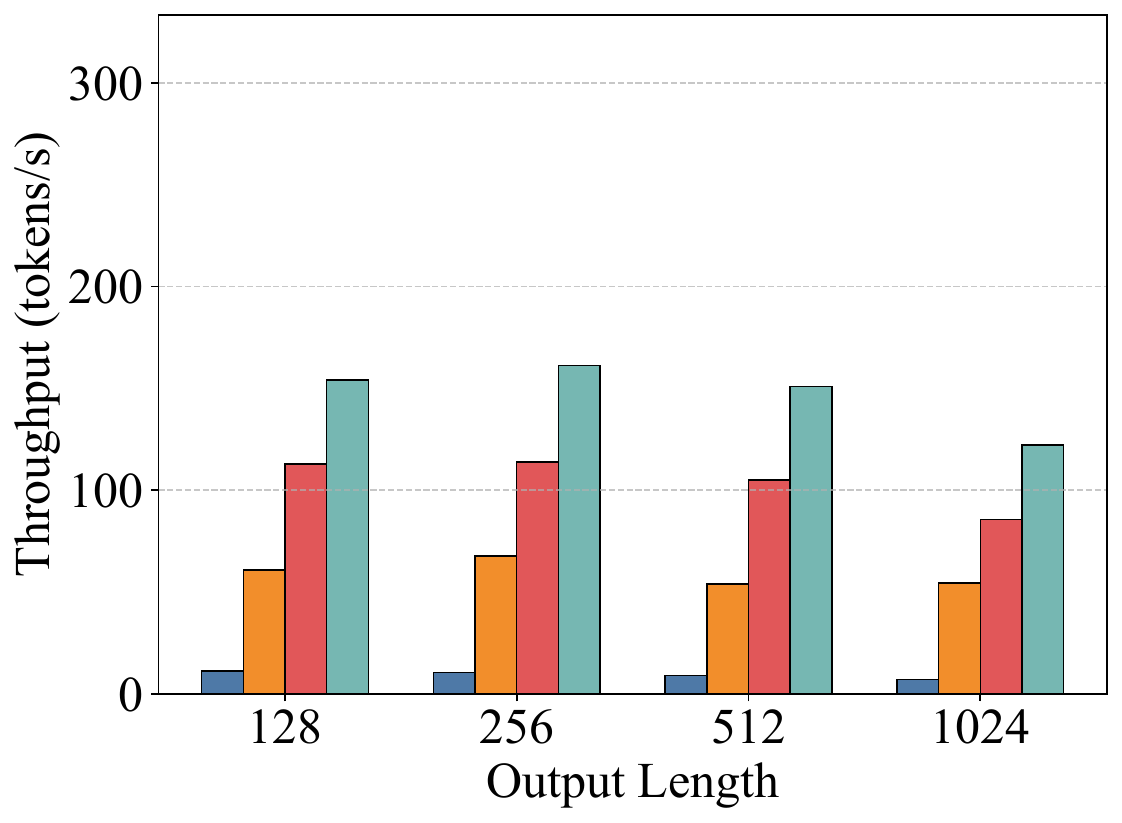}
        \caption{4090D CNN/DailyMail dataset}
    \end{subfigure}
    \vspace{-0.1in}
    \caption{End-to-end throughput comparison for A30 and 4090D across different datasets.}
    \label{fig:end2end_throughput_all}
    \vspace{-0.1in}
\end{figure*}

\begin{figure*}[t]
    \centering
    \begin{subfigure}[t]{0.24\linewidth}
        \centering
        \includegraphics[width=\linewidth]{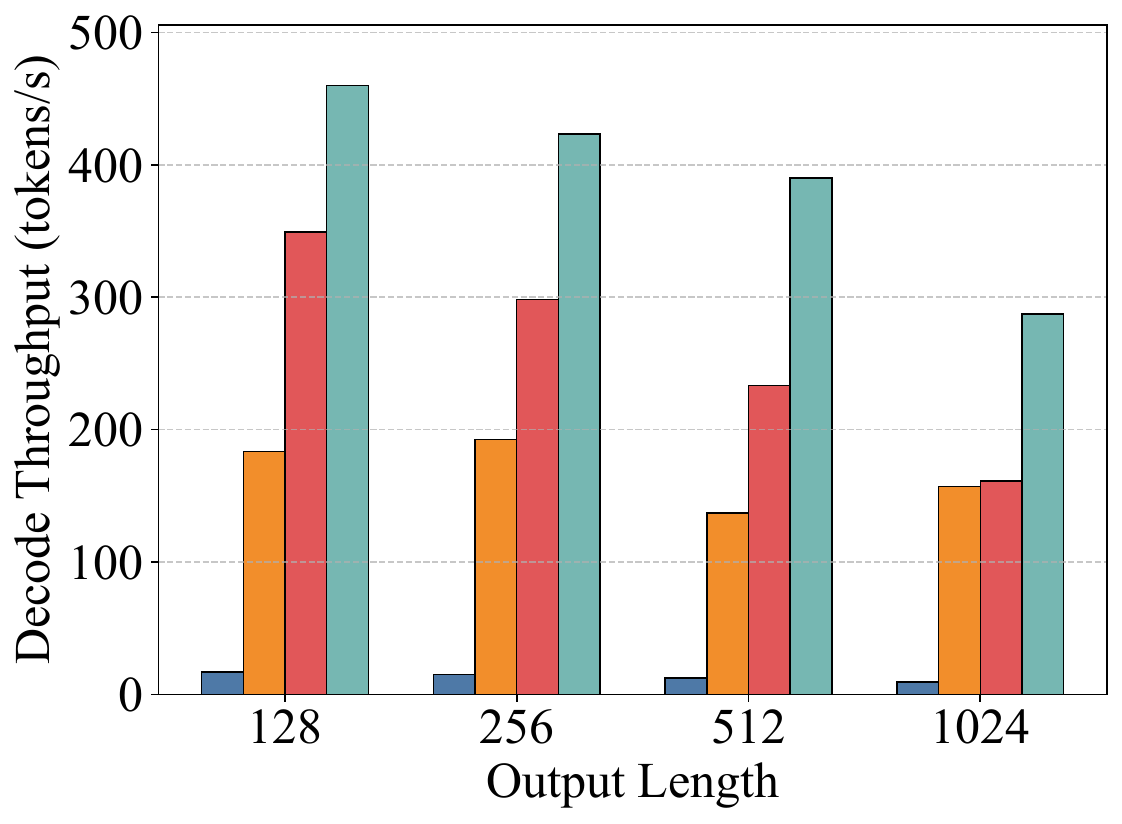}
        \caption{A30 APPS dataset}
    \end{subfigure}
    \begin{subfigure}[t]{0.24\linewidth}
        \centering
        \includegraphics[width=\linewidth]{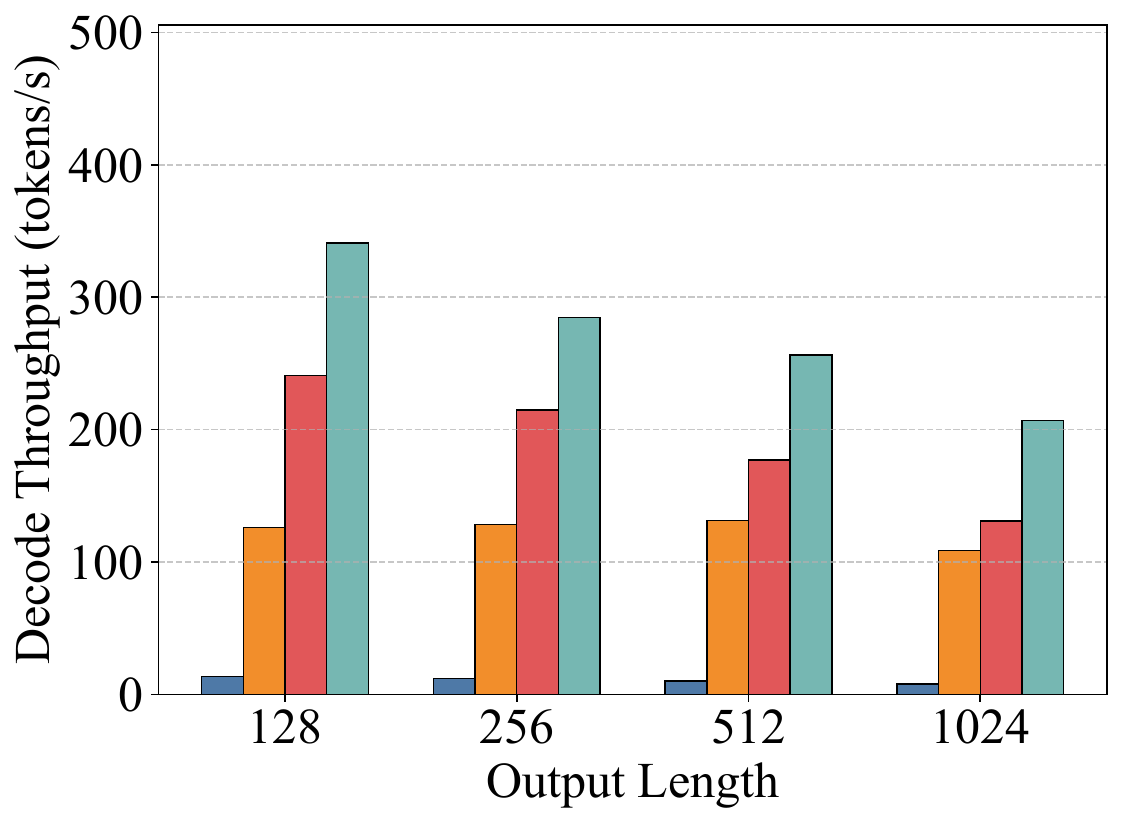}
        \caption{A30 CNN/DailyMail dataset}
    \end{subfigure}
    \begin{subfigure}[t]{0.24\linewidth}
        \centering
        \includegraphics[width=\linewidth]{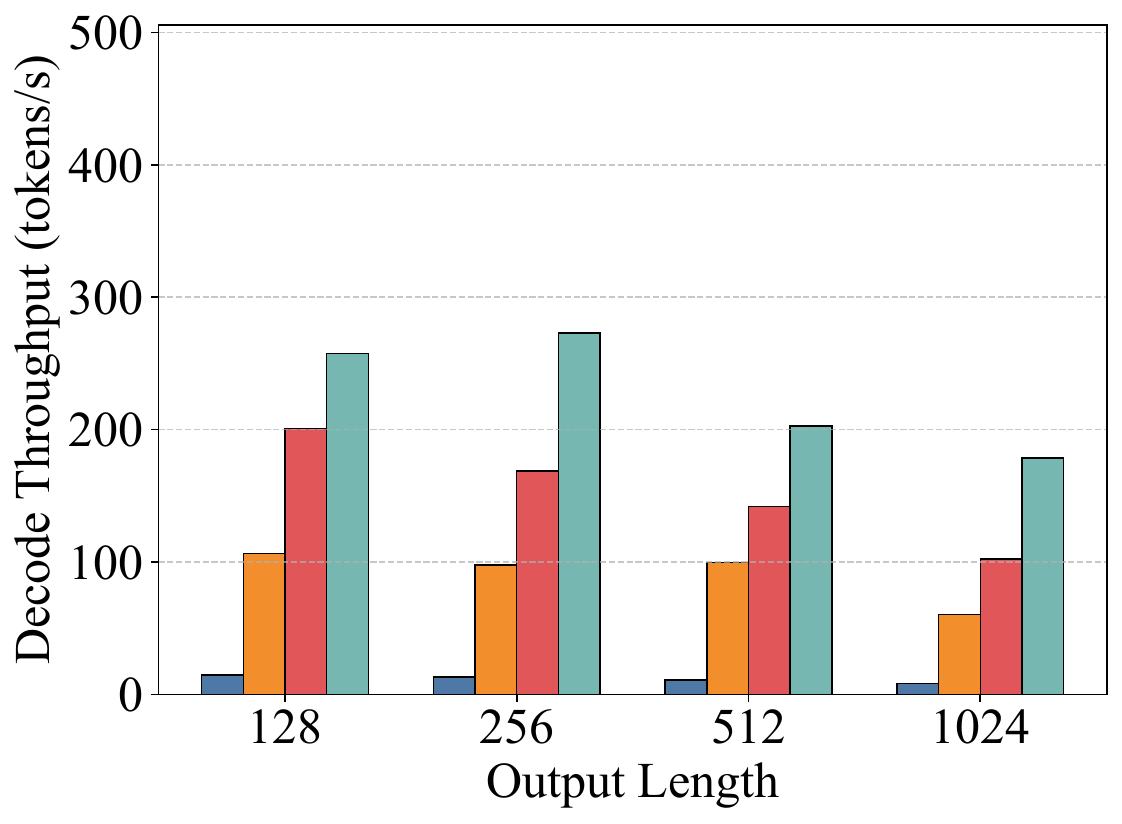}
        \caption{4090D APPS dataset}
    \end{subfigure}
    \begin{subfigure}[t]{0.24\linewidth}
        \centering
        \includegraphics[width=\linewidth]{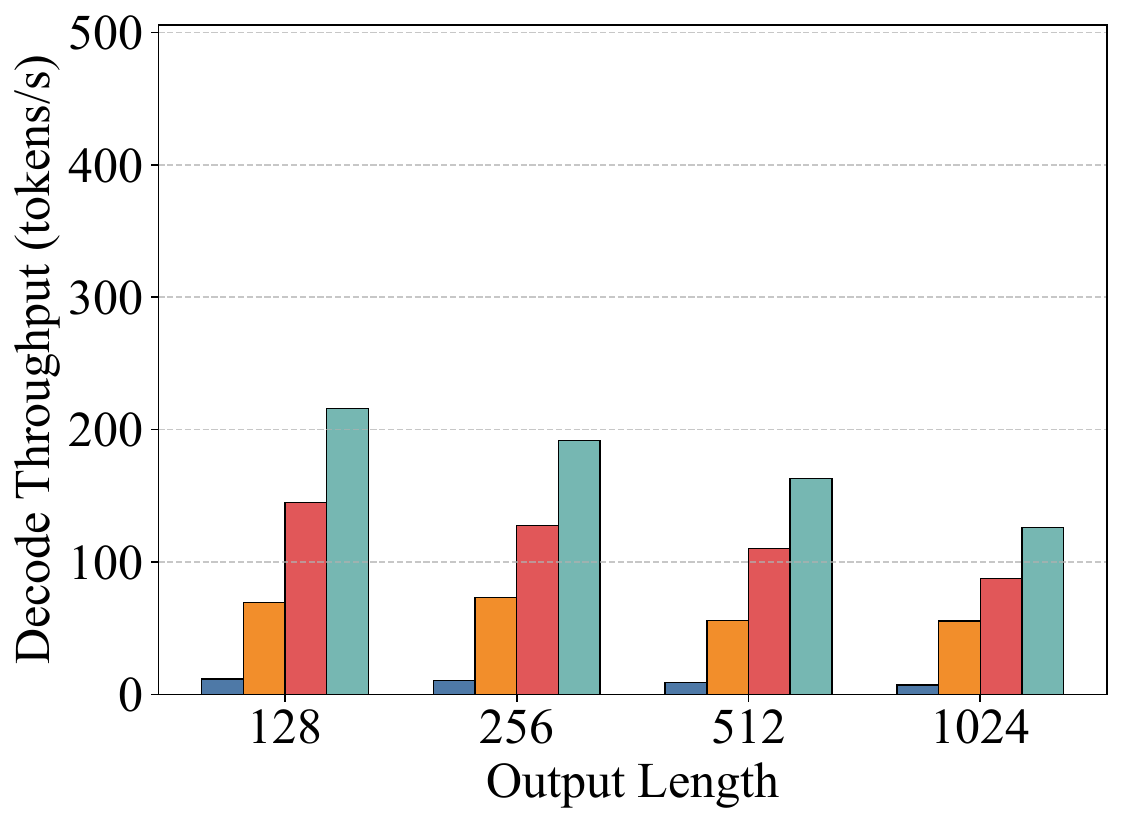}
        \caption{4090D CNN/DailyMail dataset}
    \end{subfigure}
    \vspace{-0.1in}
    \caption{End-to-end decode throughput comparison for A30 and 4090D across different datasets.}
    \vspace{-0.1in}
    \label{fig:end2end_decode_throughput_all}
\end{figure*}
\stitle{Baseline.}
We compare with two SOTA baselines designed for LLM offloading serving environment.
\emph{1). DeepSpeed-ZeRO-Inference~\cite{deepspeedinference}:} A popular LLM inference framework which support offloading model parameters to CPU memory, and dynamically load model parameters during inference. We use version 0.17.4 of DeepSpeed-ZeRO for evaluation.
\emph{2.) MoE-Lightning~\cite{cao2025moe}}: An throughput-oriented offloading serving system designed for MoE. It offloads the entire KV cache and core attention computation to the CPU.
As it achieves substantially higher throughput compared to previous offloading systems, such as FlexGen~\cite{sheng2023flexgen}, we use it as a strong baseline.
In addition we also include \name-w/o-sd, which is the variant employs \name system but disables speculative decoding, serving as a strong baseline.





\stitle{Results Overview.}
Figure~\ref{fig:end2end_throughput_all} and Figure~\ref{fig:end2end_decode_throughput_all} show the end-to-end throughput and decode throughput results, respectively, across various workloads, hardware configurations, and systems. \name outperforms DeepSpeed-ZeRO-Inference by 11.3-27.5$\times$ and average 17.7$\times$ on throughput.
We note that DeepSpeed-ZeRO-Inference suffers from small batch size due to its design of storing all requests' KV cache on GPU, which leads to low throughput.
And \name outperforms MoE-Lightning by 1.6-2.9$\times$ and average 2.1$\times$ on throughput. We also notice that \name-w/o-sd achieves higher throughput than MoE-Lightning due to superior memory management, e.g., pin memory, and better hyperparameter optimization demonstrated in Section~\ref{sec:hyperparameter_optimizer}.




\stitle{Varying dataset.}
All methods achieved higher throughput and decoding throughput on the APPS dataset compared to those on CNN/DailyMail. This is primarily attributed to the longer average input length in CNN/DailyMail, which necessitates a smaller batch size for all systems and consequently results in lower throughput. Moreover, we observe that speculative decoding (comparing \name with \name-w/o-sd) has better performance gains on the APPS dataset (1.53$\times$) compared to CNN/DailyMail (1.45$\times$). This is because code completion tasks (APPS) are generally easier to predict than summarization tasks (CNN/DailyMail).


\stitle{Varying hardware.}
Across various workloads, both \name and MoE-Lightning consistently demonstrate higher throughput on the A30 environment than on the 4090D. The key reason is the smaller CPU memory capacity of the 4090D environment, which constrains the batch size. For DeepSpeed-ZeRO, while both the A30 and 4090D have the same GPU memory, the HtoD transfer bandwidth of the 4090D is slightly lower than that of the A30, leading to comparatively lower throughput on the 4090D.

\stitle{Varying output length.}
With the increase of output length, both DeepSpeed-ZeRO-Inference, \name-w/o-sd, and \name present a decreasing trend in decode throughput due to the larger KV cache per request and decreased global batch size, while MoE-Lightning shows a relatively stable decode throughput due to its fixed optimization strategy. Regarding the throughput, both solutions exhibit a first increasing and then decreasing trend, as increasing output length leads to larger proportion of time spent on decoding and also slower decode throughput.

\begin{figure*}[t]
    \centering
    \begin{subfigure}[t]{0.33\linewidth}
        \centering
        \includegraphics[width=\linewidth]{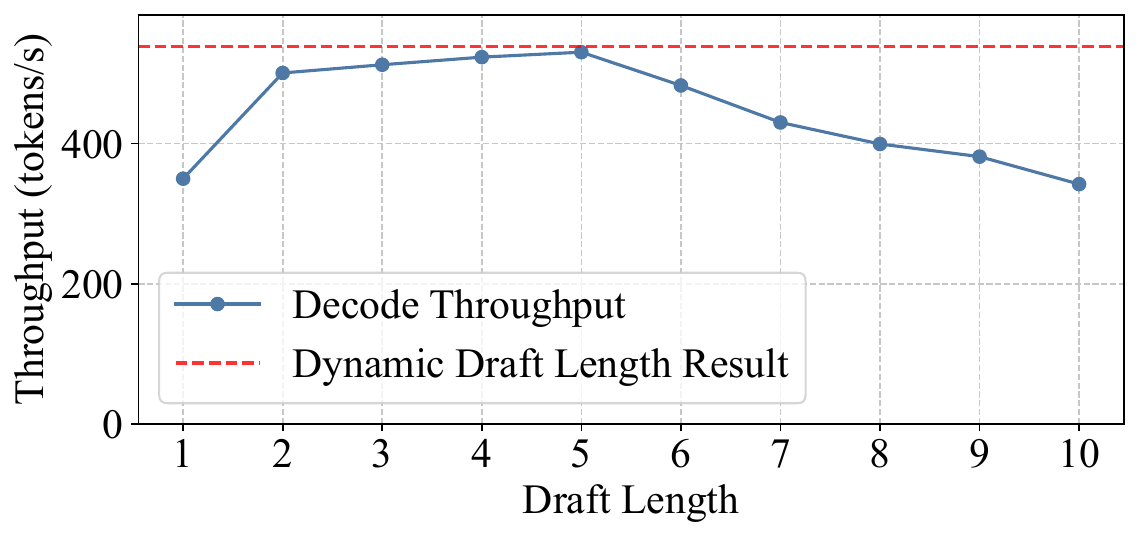}
        \caption{Throughput}
    \end{subfigure}
    \hfill
    \begin{subfigure}[t]{0.33\linewidth}
        \centering
        \includegraphics[width=\linewidth]{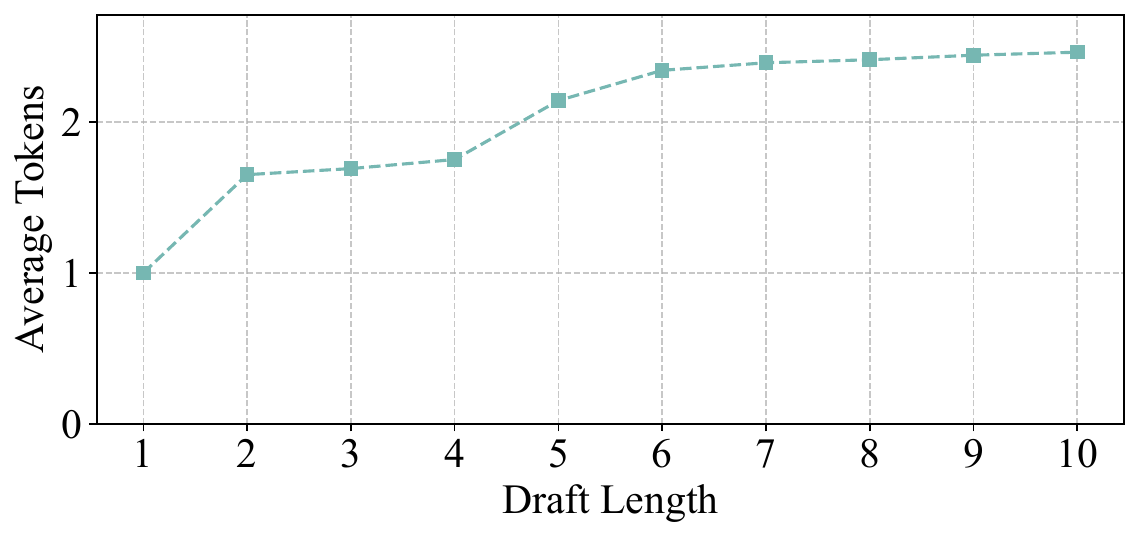}
        \caption{Average accepted tokens}
    \end{subfigure}
    \hfill
    \begin{subfigure}[t]{0.33\linewidth}
        \centering
        \includegraphics[width=\linewidth]{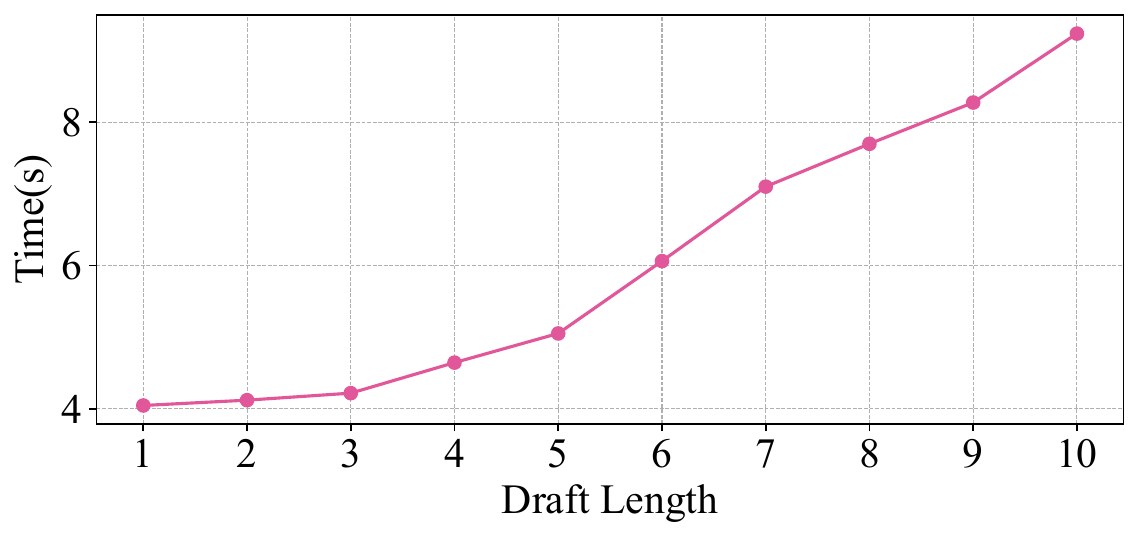}
        \caption{Iteration time} 
    \end{subfigure}
    \vspace{-0.1in}
    \caption{Impact of varying draft length.}
    \label{fig:draft_length}
\end{figure*}
\begin{figure}[t]
    \centering
    \begin{subfigure}[t]{0.48\linewidth}
        \centering
        \includegraphics[width=\linewidth]{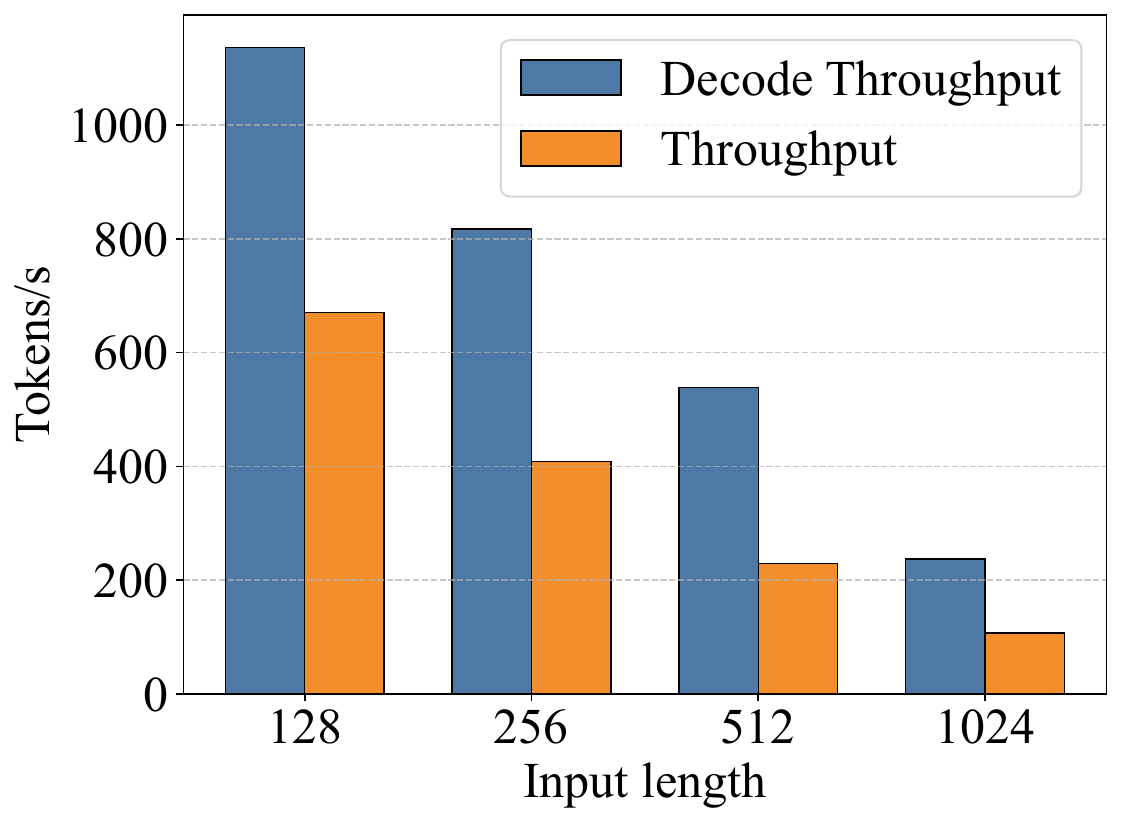}
        \caption{Fixed output length, varying input length}
    \end{subfigure}
    \hfill
    \begin{subfigure}[t]{0.48\linewidth}
        \centering
        \includegraphics[width=\linewidth]{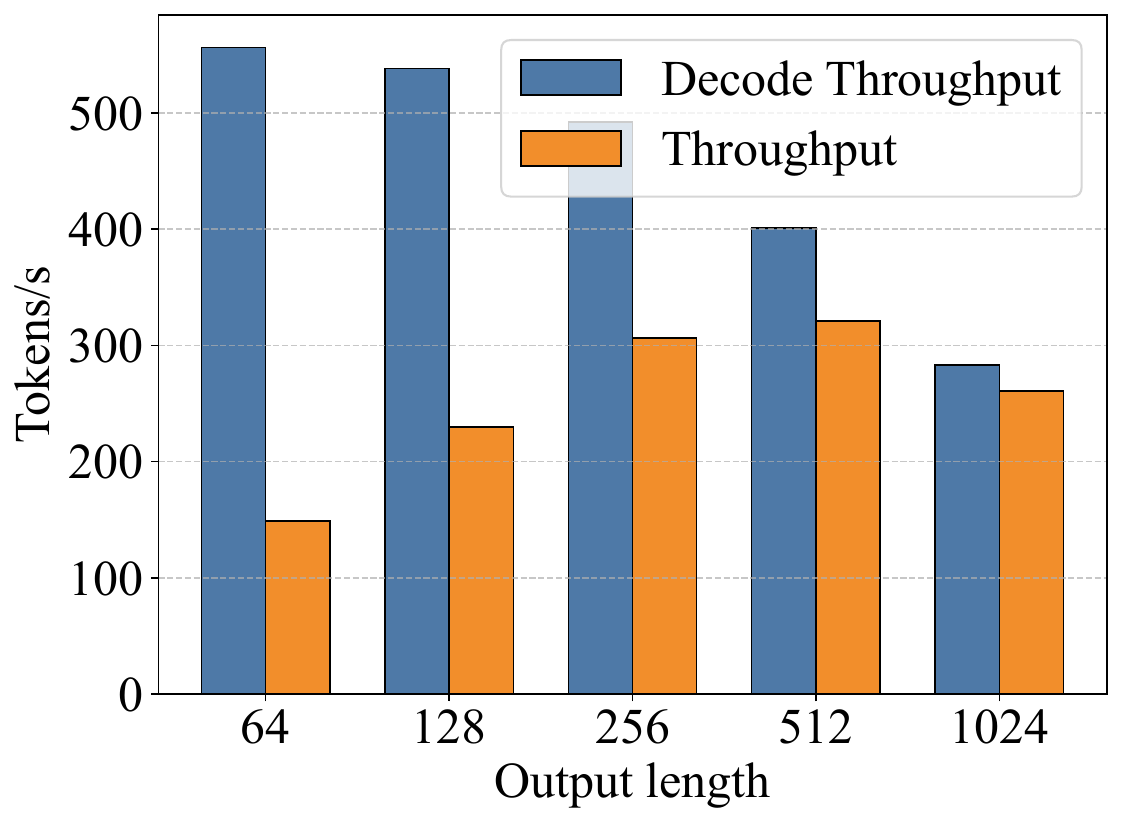}
        \caption{Fixed input length, varying output length}
    \end{subfigure}
    \vspace{-0.1in}
    \caption{Impact of varying workload.}
    \vspace{-0.1in}
    \label{fig:varying_length}
\end{figure}
\subsection{Micro-benchmark}
We conduct some micro-benchmarks to evaluate the impact of several key parameters on the system's performance, including varying hyperparameters (draft length), workload characteristics (input/output sequence length), and hardware configurations (CPU/GPU memory size). In this section, to facilitate comparisons, we by default use the APPS dataset and A30 environment as well as set the input length to 512 and output length to 128, unless otherwise specified.




\stitle{Varying draft length.}
Figure~\ref{fig:draft_length} demonstrates the impact of varying draft length from 1 to 10. Notice draft length 1 means no speculative decoding.
As the draft length increases, decode throughput shows a first increasing and then decreasing trend, as shown in Figure~\ref{fig:draft_length}(a). The initial increase in throughput is due to the fact that speculative decoding results in a higher acceptance number of draft tokens, as shown in Figure~\ref{fig:draft_length}(b), while the subsequent decrease is due to the increased iteration time, as shown in Figure~\ref{fig:draft_length}(c). We will dive deeper into the iteration time in Section~\ref{sec:delving_into_system_performance}, which shows that the CPU attention time increases with the draft length.

\name does not use a fixed draft length but dynamically adjusts it as the performance of the system determined by the current prefix length (input length + current output length).For instance, when the maximum output length is large but token generation has just begun, the draft length is set high. As more tokens are generated, the draft length is reduced. Conversely, when some requests finish generation, the draft length is increased.
The red horizontal line in Figure~\ref{fig:draft_length}(a) represents the throughput with dynamic draft length adjustment. Since the output length (128) is relatively smaller that input length (512) in this experiment, the improvement is not significant (2\%).



\stitle{Varying workload.}
As shown in Figure~\ref{fig:varying_length}, longer sequence lengths (input length + output length) result in lower decode throughput. This is because the CPU memory available for the KV cache is fixed, and longer sequences necessitate a smaller batch size.
With increasing input length, Figure~\ref{fig:varying_length}(a) shows that both throughput and decode throughput decrease.
With increasing output length, Figure~\ref{fig:varying_length}(b) shows a similar trend for decode throughput, while the overall throughput first increases and then decreases.

\begin{figure}[t]
    \centering
    \begin{subfigure}[t]{0.48\linewidth}
        \centering
        \includegraphics[width=\linewidth]{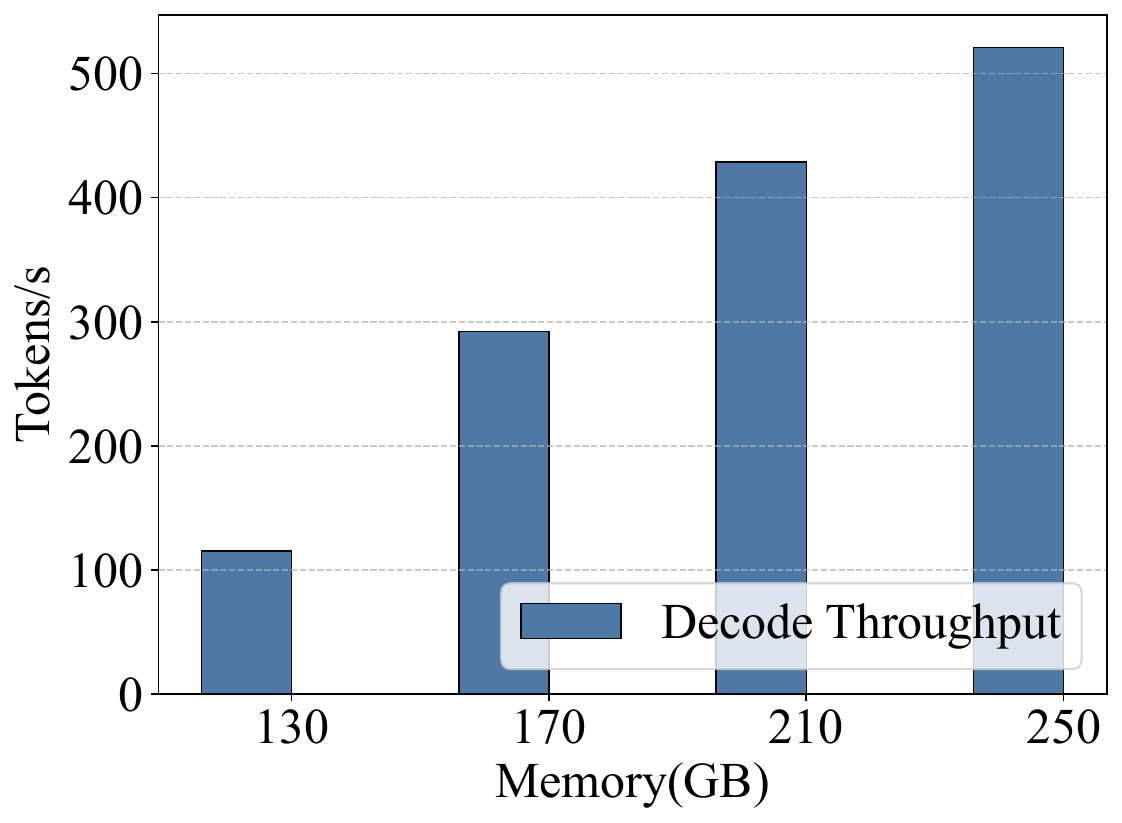}
        \caption{CPU memory}
    \end{subfigure}
    \hfill
    \begin{subfigure}[t]{0.48\linewidth}
        \centering
        \includegraphics[width=\linewidth]{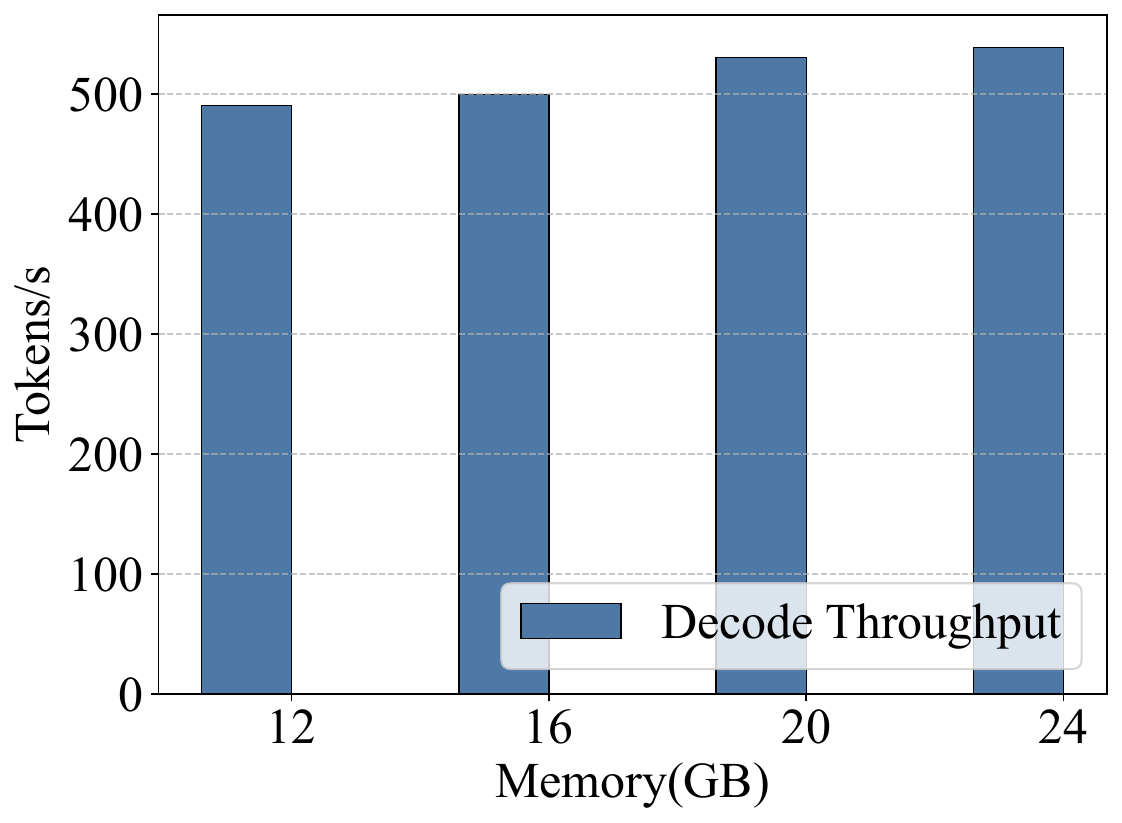}
        \caption{GPU memory}
    \end{subfigure}
    \vspace{-0.1in}
    \caption{Impact of varying CPU/GPU memory} 
    \label{fig:memory}
    \vspace{-0.2in}
\end{figure}

\stitle{Varying CPU/GPU memory.}
As shown in Figure~\ref{fig:memory}(a), increasing CPU memory capacity allows for a larger number of concurrent requests, leading to higher throughput.
Figure~\ref{fig:memory}(b) shows that increasing GPU memory capacity has a limited impact on throughput. This is because the GPU memory is relatively small compared to the CPU memory, thereby the incremental benefits of cache size, i.e., GPU memory, are marginal.

\subsection{Delving into System Performance}
\label{sec:delving_into_system_performance}
We first evaluate the profiling-based estimator's accuracy which directly affects the hyperparameter optimization, and then we analyze the iteration time breakdown to understand the system performance.

\stitle{Estimation Accuracy.}
With a fixed sequence length of 512, an output length of 128, and a draft token count of 5, Table \ref{tab:performance_comparison} presents the average iteration execution time, the time consumed by individual components, and the time estimated by the performance estimator. The overall estimation error is within 10\%.

\begin{table}[htbp]
    \centering
    \caption{Performance Comparison: Actual vs. Estimated}
    \label{tab:performance_comparison}
    \begin{tabular}{lrrr}
        \hline
        \hline
                              & \textbf{Actual} & \textbf{Estimated} & \textbf{Error} \\ \hline
        \textbf{Target model} & \textbf{4.39s}  & \textbf{4.03s}     & 8.2\%          \\
        CPU Attention         & 4.29s           & 3.88s              & 10.6\%         \\
        GPU MoE               & 3.53s           & 3.17s              & 10.2\%         \\
        HtoD Transfer         & 3.70s           & 3.55s              & 4.1\%          \\ \hline
        \textbf{Draft model}  & \textbf{0.56s}  & \textbf{0.41s}     & 26.8\%         \\
        GPU Part              & 0.42s           & 0.35s              & 16.7\%         \\
        CPU Part              & 0.54s           & 0.41s              & 24.1\%         \\ \hline
        \textbf{Others}       & \textbf{0.097s} & \textbf{0.12s}     & 23.7\%         \\ \hline
        \textbf{Iteration}    & \textbf{5.05 s} & \textbf{4.56 s}    & 9.7\%          \\ \hline
    \end{tabular}
\end{table}

\begin{figure}[htbp]
    \centering
    \includegraphics[width=0.9\linewidth]{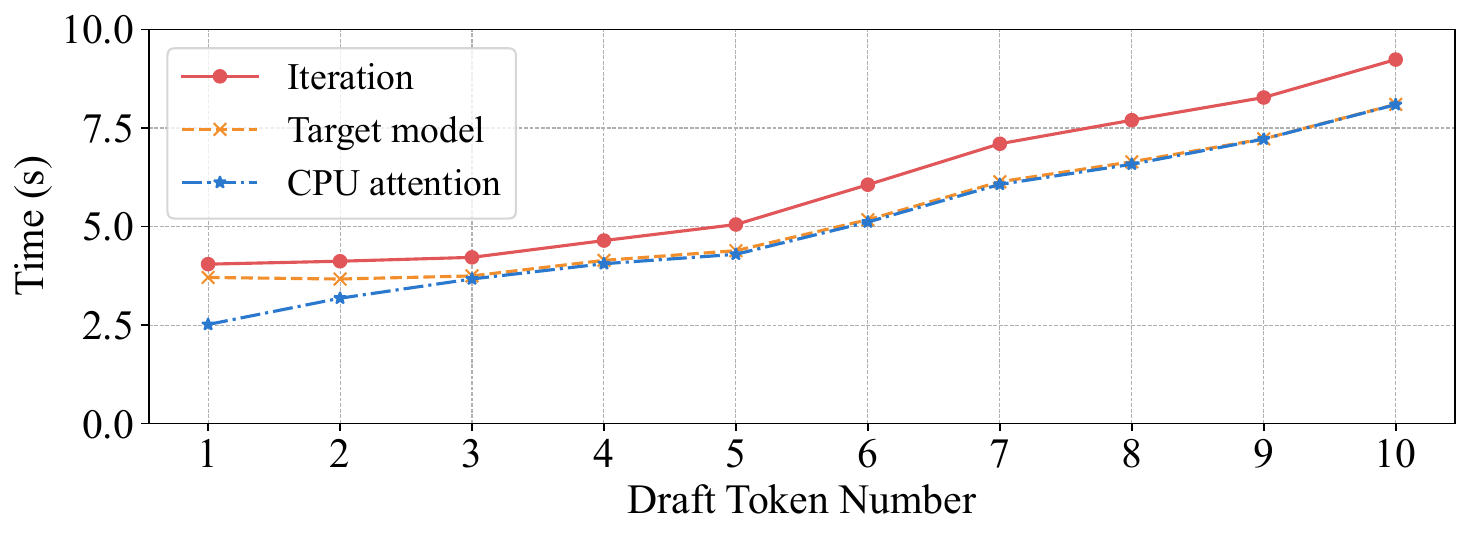}
    \vspace{-0.12in}
    \caption{Iteration time breakdown.} 
    \label{fig:iteration_breakdown}
    \vspace{-0.2in}
\end{figure}

\stitle{Iteration Breakdown.}
For a more granular analysis of system performance, as shown in Figure~\ref{fig:iteration_breakdown}, we break down the execution time of a single iteration. We observed that as the number of draft tokens increases, the proportion of time spent on CPU attention within the target model grows, gradually becoming a performance bottleneck for the target model. Furthermore, the overhead associated with draft generation also increases with a higher number of draft tokens.

\section{Conclusion}
In this paper, we present \name, the first MoE offloading system that leverages speculative decoding to increase hardware utilization and improve throughput. \name employs speculative decoding to enhance both CPU and GPU utilization, thereby improving throughput.
To fully exploit the potential of speculative decoding in offloading scenarios, \name carefully orchestrates the execution of the target and draft models, designing dedicated CPU-based chunked attention operators,
memory-conscious draft generation, and hyperparameter optimization.
The experimental results demonstrate that \name achieves a decode throughput of 2.5$\times$ superior to the state-of-the-art MoE offloading serving system, MoE-Lightning.





\bibliographystyle{plain}
\bibliography{main.bib}

\end{document}

%% file: command.tex
\makeatletter
\newcommand{\removelatexerror}{\let\@latex@error\@gobble}
\makeatother

\newcommand{\eat}[1]{}

\long\def\comment#1{}

\newcommand{\stitle}[1]{\vspace{.5ex} \noindent{{\bf #1}}}

\newcommand{\name}{SpecMoEOff\xspace}